\documentclass[aps,prb, floatfix,notitlepage,nofootinbib,twocolumn,superscriptaddress,reprint]{revtex4-2}
\usepackage{amsmath}
\usepackage{amssymb}
\usepackage{mathtools}
\usepackage{graphicx}
\usepackage[tight]{subfigure}
\usepackage{enumitem}
\usepackage{soul}
\usepackage{pifont}
\usepackage{xspace}
\usepackage{makecell}
\usepackage{lipsum}
\usepackage{xcolor}

\usepackage{multirow}

\newcommand{\pcr}{\fontfamily{pcr}\selectfont}

\usepackage[english]{babel}
\makeatletter

\def\bbl@set@language#1{%
  \edef\languagename{%
    \ifnum\escapechar=\expandafter`\string#1\@empty
    \else\string#1\@empty\fi}%

  \@ifundefined{babel@language@alias@\languagename}{}{%
    \edef\languagename{\@nameuse{babel@language@alias@\languagename}}%
  }%
  \select@language{\languagename}%
  \expandafter\ifx\csname date\languagename\endcsname\relax\else
    \if@filesw
      \protected@write\@auxout{}{\string\select@language{\languagename}}%
      \bbl@for\bbl@tempa\BabelContentsFiles{%
        \addtocontents{\bbl@tempa}{\xstring\select@language{\languagename}}}%
      \bbl@usehooks{write}{}%
    \fi
  \fi}

\newcommand{\DeclareLanguageAlias}[2]{%
  \global\@namedef{babel@language@alias@#1}{#2}%
}
\makeatother
\DeclareLanguageAlias{en}{english}

\usepackage{bm}
\renewcommand{\vec}[1]{\boldsymbol{\mathbf{#1}}}

\def\P{\mathbb P}

\def\diffd{\mathrm{d}}
\AtBeginDocument{}

\DeclareMathOperator\proba{\mathbb P}
\newenvironment{spmatrix}{\big(\begin{smallmatrix}}{\end{smallmatrix}\big)}

\usepackage[pdfusetitle,colorlinks=true,citecolor=blue,linkcolor=magenta]{hyperref}
\usepackage{nameref}
\usepackage[capitalise,compress]{cleveref}
\usepackage{braket}

\graphicspath{{./}{./images/}}

\def\bea{\begin{equation}\begin{aligned}}
\def\eea{\end{aligned}\end{equation}}
\usepackage[normalem]{ulem}

\long\def\EB#1{{\color{purple}\ifmmode\text{\scriptsize[#1]}\else\footnotesize{[EB: #1]}\fi}}

\usepackage[ruled]{algorithm2e}
\begin{document}

\title{Operator Spreading, Duality, and the Noisy Long-Range FKPP Equation}
\author{Tianci Zhou}
\affiliation{Department of Physics, Virginia Tech, Blacksburg, Virginia 24061, USA}
\author{\'Eric Brunet}
\affiliation{Sorbonne Universit\'e, Laboratoire de Physique de l'\'Ecole Normale Sup\'erieure}
\affiliation{ENS, Universit\'e PSL, CNRS}
\affiliation{Universit\'e Paris Cit\'e, Paris, France}
\author{Xiaolin Qi}
\affiliation{Department of Physics, Virginia Tech, Blacksburg, Virginia 24061, USA}
 
\date{\today}

\begin{abstract}
Operator spreading provides a new characterization of quantum chaos beyond the semi-classical limit. There are two complementary views of how the characteristic size of an operator, also known as the butterfly light cone, grows under chaotic quantum time evolution: A discrete stochastic population dynamics or a stochastic reaction-diffusion equation in the continuum. When the interaction decays as a power function of distance, the discrete population dynamics model features superlinear butterfly light cones with stretched exponential or power-law scaling. Its continuum counterpart, a noisy long-range Fisher-Kolmogorov-Petrovsky-Piscunov (FKPP) equation, remains less understood. We use a mathematical duality to demonstrate their equivalence through an intermediate model, which replaces the hard local population limit by an equilibrium population. Through an algorithm with no finite size effect, we demonstrate numerically remarkable agreements in their light cone scalings.  
\end{abstract}

\maketitle

\begin{figure}
    \centering
    \includegraphics[width=1\linewidth]{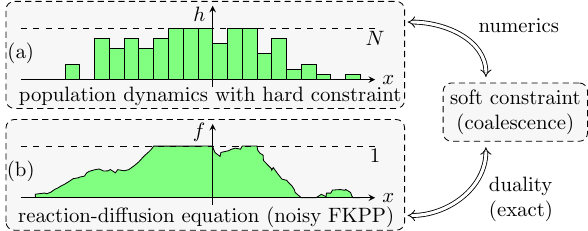}
    \caption{Two perspectives of understanding operator spreading: (a) population dynamics and (b) reaction-diffusion equation. We establish their equivalence through an intermediary population dynamics model with a soft constraint.} 
    \label{fig:schematic}
\end{figure}

{\it Introduction.} Recent research on quantum chaos beyond semi-classical limits has been shaped by the concept of operator spreading. The idea is inspired by the Lyapunov behaviors of an out-of-time ordered correlator (OTOC)\cite{larkin_quasiclassical_1969,shenker_black_2014,roberts_diagnosing_2015,kitaev2015,maldacena_bound_2015} which can be quantitatively characterized by the growth of an operator's support in the Heisenberg picture\cite{nahum_operator_2018,von_keyserlingk_operator_2018,zhou_operator_2020,xu_locality_2018,chen_quantum_2018,zhou_operator_2018,roberts_operator_2018,qi_quantum_2018,de_stochastic_2024}. There are two complementary perspectives to describe the growth process. First, it can be viewed as a stochastic process of a non-negative integer variable $h$ with uniform upper bound $N$ on a lattice (Fig.~\ref{fig:schematic}(a)). In a qubit system, for instance, $h$ typically represents the number of Pauli operators at a given site (commonly referred to as the Pauli weight). Notably,  in the OTOCs of various random quantum circuits\cite{lashkari_towards_2013,nahum_operator_2018,von_keyserlingk_operator_2018,zhou_operator_2020,xu_locality_2018,chen_quantum_2018,zhou_operator_2018,roberts_operator_2018,qi_quantum_2018,agarwal_emergent_2021,erdos_phase_2014,lucas_quantum_2019,vardhan_entanglement_2024}, the growth of $h$ in space has a random walk behavior and can be interpreted as a form of population dynamics. The second perspective describes the dynamics as a stochastic reaction-diffusion equation in spatial continuum, where the field is a continuous approximation of $f=\frac{h}{N}$(Fig.~\ref{fig:schematic}(b)). While the mean-field limit ($N \rightarrow \infty$) is deterministic, the large-$N$ expansion introduces noise that can quantitatively affect the spreading. In Refs.~\cite{aleiner_microscopic_2016,aron_kinetics_2023,aron_traveling_2023} ,  a noiseless Fisher-Kolmogorov-Pestrovsky-Piscunov (FKPP) equation\cite{kolmogorov_investigation_1937,fisher_wave_1937,ablowitz_explicit_1979,brunet_aspects_2016}  of OTOC is derived in strongly interacting fermionic systems, where the resulting wavefront characterizes the propagation of chaos. For both views, the assumption of quantum chaos justifies the use of stochastic interactions in the quantum model, leading to effective classical Markovian dynamics. In this paper, we investigate the equivalence of these two complementary views through the tool of duality. 

To set the stage, let us illustrate the two perspectives using a qubit system example. Suppose $H$ is a Hamiltonian consisting of locally interacting terms and $X_n$ is a Pauli operator at position $n$. The Heisenberg evolution $i[H, X_n]\,\diffd t$ at the initial time increases the operator's support, as non-commutive terms with $X_n$ generate longer Pauli strings (i.e., larger Pauli weight) on top of $X_n$.  Repeated application of commutator over time leads to operator growth. In a quantum dot model, there can be $N > 1$ spins on each site. Under the assumption of chaos, such operator growth can be mapped to a population growth model\cite{xu_locality_2018,zhou_operator_2018,zhou_operator_2020,qi_quantum_2018} with a birth-at-distance process and a population limit $N$ on each site. The mapping becomes exact in a minimal model of Brownian quantum circuit\cite{lashkari_towards_2013,Xu_Swingle_2018,agarwal_emergent_2021,bentsen_measurement-induced_2021,erdos_phase_2014,lucas_quantum_2019,vardhan_entanglement_2024}. 
In the mean-field limit($N \rightarrow \infty$), the normalized population $f = \frac{h}{N}$ becomes a continuous variable in $[0,1]$. The birth-at-distance process maps to reaction and diffusion of $f$, and the evolution is deterministic. For large but finite $N$, the fluctuation and discreteness of $f$  bring in a noise correction of order $\frac{1}{\sqrt{N}}$ \cite{brunet_aspects_2016} .
Finally, by taking the continuum limit in space, the dynamics are governed by the stochastic FKPP equation. 
Consistency requires that both perspectives yield the same statistics for a finite $N$. Here we focus on the statistical properties of chaos propagation, known as the butterfly light cone\cite{roberts_lieb-robinson_2016}. It is the region where the OTOC attains an appreciable value and the variable $h$ or $f$ (depending on the perspectives) reaches a threshold in the two perspectives. 

In the local short-range interaction case, consistency is known to be satisfied\cite{nahum_operator_2018,von_keyserlingk_operator_2018,brunet_aspects_2016,xu_locality_2018,zhou_operator_2018}.  In one spatial dimension, the stochastic FKPP equation produces a linear propagating wave with a diffusive front\cite{brunet_aspects_2016}, which agrees with the population dynamics where the interface of the high/low population regimes undergoes a drifted random walk\cite{nahum_operator_2018,von_keyserlingk_operator_2018}. In the long-range interacting systems, the connection between the two views is less clear.

In this work, we relate the two perspectives through a duality relation (Fig.~\ref{fig:schematic}). Such a duality was already known\cite{doering_interacting_2003} for the local FKPP equation.

The duality is a mathematical statement that certain combined correlators of the population dynamics and stochastic equation are time-independent so that the dynamics of one constrains the other. As a result, the light cone scalings of the two models are the same. On the population dynamics side\cite{zhou_operator_2020,hallatschek_acceleration_2014,qi_quantum_2018}, there is a ``population limit" -- a hard constraint -- imposed by the maximal number $N$ of spin-$\frac{1}{2}$s at each site. We are not able to implement this constraint exactly in the dual model. Instead, we introduce a coalescence process in the population dynamics, so that the {\it equilibrium population} is $N$. We call this a (dynamical) soft constraint. We seek the chain of equivalence given by Fig.~\ref{fig:schematic}. 

We perform comprehensive numerical simulations of both the hard and soft constraint models in 1D. To preserve the long tail of the power law and remove finite size effects, the program tracks the variable $h$ across a window of $10^9$ sites and uses a sparse representation to record particles beyond that range. We find that the coalescence model has a remarkable agreement with the scaling predictions of the hard constraint model. Since the coalescence model is exactly dual to the stochastic long-range FKPP equation, we conclude that the two perspectives share identical butterfly light cone scaling (for $\alpha > 0.5$ in 1D), and provide consistent descriptions of how the operator spreads in generic long-range interacting quantum systems(Fig.~\ref{fig:schematic}(a)). Analytically, we perform a third order perturbative analysis about the scaling function close to the critical point $\alpha = 1$. The match with the short and median time scalings provides practical benchmarks when asymptotic scaling is not achievable in either numerics or experiments. 

{\it Duality and coalescence.}
We begin by precisely defining the population dynamics models we study. Operator spreading in a qubit system is generally defined on a $d$-dimensional lattice whose sites host an integer height variable $h( \vec{x} , t) $ at position $\vec x$ and time $t$ representing the number of Paulis at each site, or equivalently the local population size in population dynamics. When the evolution is chaotic, it generates a Markov process of height-increasing events. Specifically, a site at position $\vec x$ can increase $h(\vec{y}, t)$ at a site $\vec y$ by $1$ with a rate proportional to $G( \|\vec{x} - \vec{y} \|) h(\vec x,t)(1-\frac{1}{N}h(\vec y,t))$. In a chaotic system, coherent contributions dephase and only their fluctuations accumulate; the kernel $G(r)$ decays as $\frac{1}{r^{2\alpha}}$ when the quantum interactions decay as $\frac{1}{r^{\alpha}}$. We thus require $\alpha > \frac{d}{2}$ for the total rate to be finite. The factor of  $1 -\frac{1}{N} h(\vec{y},t )$ in the rate limits the height to be at most $N$. Note that when a height-increasing event occurs, the population at $\vec y$ is increased, but the population at site $\vec x$ remains  \emph{unchanged}: it is a ``birth at distance'', not a ``jump''. We call it the hard constraint model. 
\begin{figure}
    \centering
    \includegraphics[width=1\linewidth]{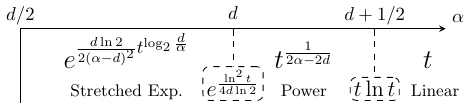}
    \caption{Scalings functions of the light cones for the hard constraint model in $d$-dimension.}
    \label{fig:model_rate_phase}
\end{figure}

In general, a duality relates the stochastic dynamics of $h(\vec{x},t)$ with another continuous stochastic field $f(\vec{x}, t) \in [0, 1]$, whose time evolution is designed so that, for any $t$, and any choice of initial conditions $f(\vec x,0)$ and $h(\vec x,0)$, the combined correlator
\begin{equation}
\label{eq:martingle}
    \langle \prod_{\vec{x}} (1- f( \vec{x}, \tau ))^{h( \vec{x}, t- \tau ) }\rangle_{f, h} 
\end{equation}
remains independent of $\tau$ in a double ensemble average over $f$ and $h$.

The time independence of Eq.~\ref{eq:martingle} directly links the light cones of the two models. By duality, $\tau = 0$ and $\tau = t$ are related through
\begin{equation}
\label{eq:duality_0_tau}
\begin{aligned}
    \langle \prod_{\vec{x}} (1- f( \vec{x}, 0 ))^{h( \vec{x}, t ) }\rangle_{h} 
     = \langle \prod_{\vec{x}} (1- f( \vec{x}, t ))^{h( \vec{x}, 0 ) }\rangle_{f} .
\end{aligned}
\end{equation}
Now consider an initial condition where both $f$ and $g$ have delta function profiles, with the two deltas separated by a distance $L$: $f( \vec{x}, 0 ) =\frac12 \delta_{\vec x}^{L\vec e_x}$ , $h( \vec{x} , 0 ) = \delta_{\vec x}^{\vec 0}$. Then the duality relation between $0$ and $t$ (Eq.~\ref{eq:duality_0_tau}) simplifies to
\begin{equation}
    \big\langle 2^{-h( L\vec{e}_x, t ) }\big\rangle_h = 1 - \big\langle f( \vec{0}, t )\big\rangle_f.
\end{equation}
The relation indicates that $h( L\vec{e}_x, t ) \simeq 0$ is equivalent to $f(\vec{0}, t ) \simeq 0$. In other words, if the operator spreading has (not) reached a distance $L$ from its initial peak, the wave of the process $f$  must also have (not) traveled a distance $L$ from its initial peak. Hence, the light cone scalings of the two models are identical. 

It now remains to determine the dynamics of $f$ so that the duality holds. In the construction, each event in $h$ corresponds to a term in the partial differential equation (PDE) satisfied by $f$. If the rates were $G(\| \vec{x} -\vec {y} \| ) h(\vec{x}) $,  meaning the factor of $1 - \frac{1}{N}h(\vec{y} , t )$ were ignored, they would be collectively dual to a superdiffusion and a logistic term $f(1-f)$.  However, we are unable to construct the dual term that directly enforces the maximal height limit required by  $1 - \frac{1}{N} h(\vec{y} , t )$ factor in the rate. 

To proceed, we propose to replace the hard constraint with a soft constraint induced by a pairwise coalescence process $A + A \rightarrow A$ at each site. More specifically, in the new model, birth occurs at position $\vec y$ from position $\vec x$ with a rate $h(\vec x,t) G(\|\vec y-\vec x\|)$, without the $1-\frac1Nh(\vec y,t)$ factor constraining the population size not to exceed $N$. To compensate, independently on each site, the population size $h$ decreases by one unit with rate $\beta \frac{1}{2} h(h - 1)$.  This model will reach equilibrium as the coalescence rate increases faster than the growth rate for large $h$. We call it the soft constraint model. 

The soft constraint imposed by the coalescence is exactly dual to a stochastic term for the equation on the lattice. In the continuum limit in 1D,  it leads to a PDE similar to the stochastic FKPP equation, which has the following form with rescaled parameters $\tilde{D}$, $\tilde{\lambda}$ and $\tilde{\beta}$  \cite{SM}: 
\begin{equation}
\label{eq:noisy_fkpp}
    \partial_t f
     = -(1 - f) \tilde D(-\Delta)^{\frac{\mu}{2}} f + \tilde\lambda f ( 1 - f ) + \sqrt{\tilde\beta f (1 - f)} \,\eta . 
\end{equation}
where $\mu = 2\alpha - 1$, $\eta(x, t)$ is a space-time white noise. Note that the dual function $f(x,t)$ in continuum cannot be identified with the hydrodynamic limit for $\frac{h}{N}$ when $N$ is large. The duality holds for all positive integer $N$ even as small as 1. However, the equation for the hydrodynamic limit of $\frac{h}{N}$ is expected to be very similar. 

We will use the soft constraint model as an intermediary (Fig.~\ref{fig:schematic}). The exact duality argument above shows its equivalence to the stochastic FKPP equation. Evidence for the equivalence between the soft and hard constraints can thus unify the two perspectives in the introduction. 

{\it Iterative scaling argument}. The leading order light cone scalings of the $N = 1$ hard constraint model are known exactly\cite{hallatschek_acceleration_2014,zhou_operator_2020,chatterjee_multiple_2013}, and Fig.~\ref{fig:model_rate_phase} lists the scaling functions in $d$-dimension. To summarize, the model exhibits stretched exponential  ($d > \alpha > \frac{d}{2}$ ), power-law ($ d+ \frac{1}{2} \ge \alpha  > d$) and linear ($\alpha > d + \frac{1}{2}$) light cones as $\alpha$ increases. The superlinear light cone, especially the stretched exponential type, poses a significant challenge for numerical verification.

\begin{figure}[h]
    \centering
    \includegraphics[width=0.33\columnwidth]{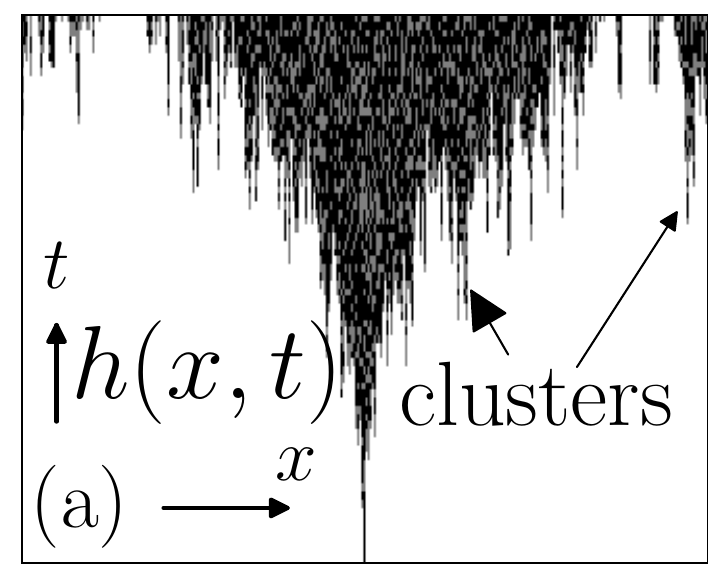}
    \includegraphics[width=0.65\columnwidth]{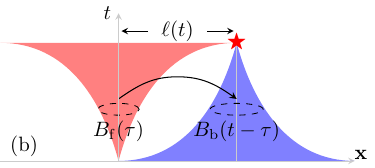}
    
    \caption{Butterfly light cones.  (a) A greymap of a $h(x, t)$ instance for the coalescence model with $\alpha = 1$ and equilibrium height $\approx 2$ for $x \in [-10^5, 10^5]$ and $t \in [0,40]$. Each pixel is white for $h=0$, grey for $h=1$ , and black for $h\ge2$. (b) A heuristic picture to derive the consistency equation of the light cone.}
    \label{fig:lc_and_ilc}
\end{figure}

The light cone scalings were obtained in Ref.~\cite{hallatschek_acceleration_2014} from an iterative scaling argument that describes how light cones form through cluster seeding and merging, see Fig.~\ref{fig:lc_and_ilc}(a).
The long-range part of birth-at-distance can seed clusters outside a linear light cone, which can subsequently merge with the main cluster to create a superlinear growth. We illustrate the (generically superlinear) light cone in Fig.~\ref{fig:lc_and_ilc}(b), with the star marking its typical radius $\ell(t)$ --- this is the size of the light cone. For the star to be occupied at time $t$, a seed must exist in the past light cone of the star, which is depicted as an inverted funnel in Fig.~\ref{fig:lc_and_ilc}(b). Ref.~\cite{hallatschek_acceleration_2014} argued that there are  $\mathcal{O}(1)$ jumps between the forward and backward light cones (more than $\mathcal{O}(1)$ jumps will create a larger light cone). This leads to the integral equation:
\begin{equation}
\label{eq:iter_equ}
    \int_{0}^{t} {\rm d}\tau \int_{B_{\rm b}(t-\tau)} \,{\rm d}\vec y \int_{B_{\rm f}(\tau)} \, {\rm d}\vec x \,  G( \|\vec{x} - \vec{y}\|) \sim 1
\end{equation}
where $B_\text f(\tau)$ is the ball centered on the origin of radius $\ell(\tau)$, and $B_\text{b}(t-\tau)$ is the ball centered on a point at distance $\ell(t)$ from the origin and of radius $\ell(t-\tau)$, see Fig.~\ref{fig:lc_and_ilc}(b). 

\begin{figure*}
\centering
 \includegraphics[width=0.323\textwidth]{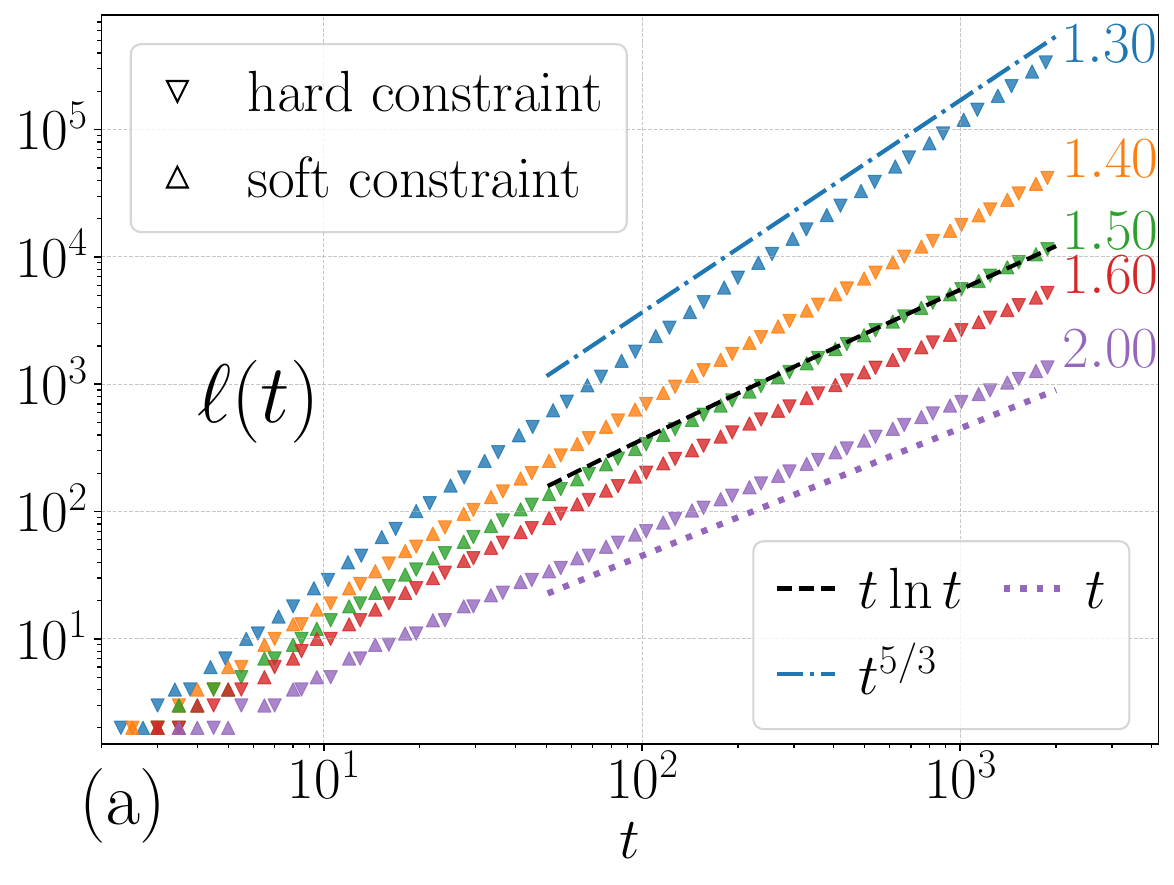}
 \includegraphics[width=0.325\textwidth]{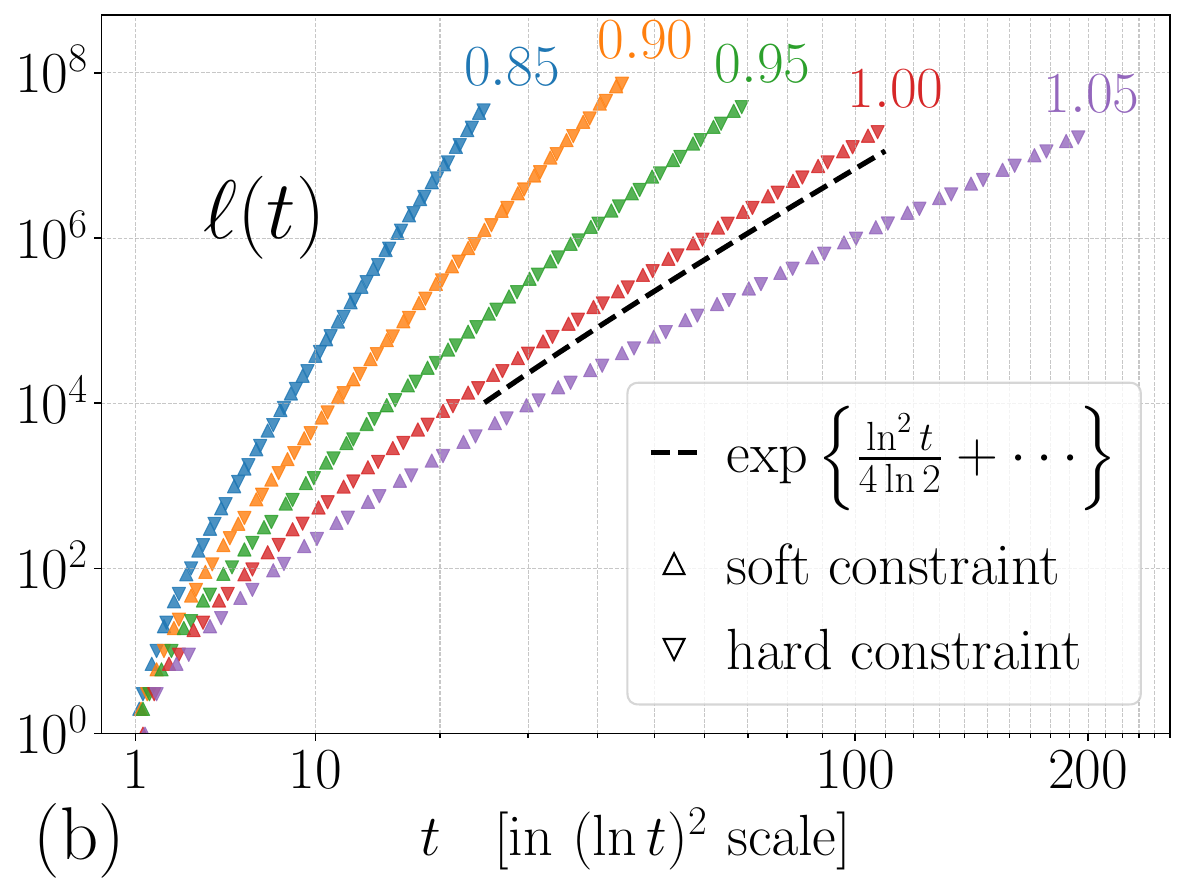}
  \includegraphics[width=0.33\textwidth]{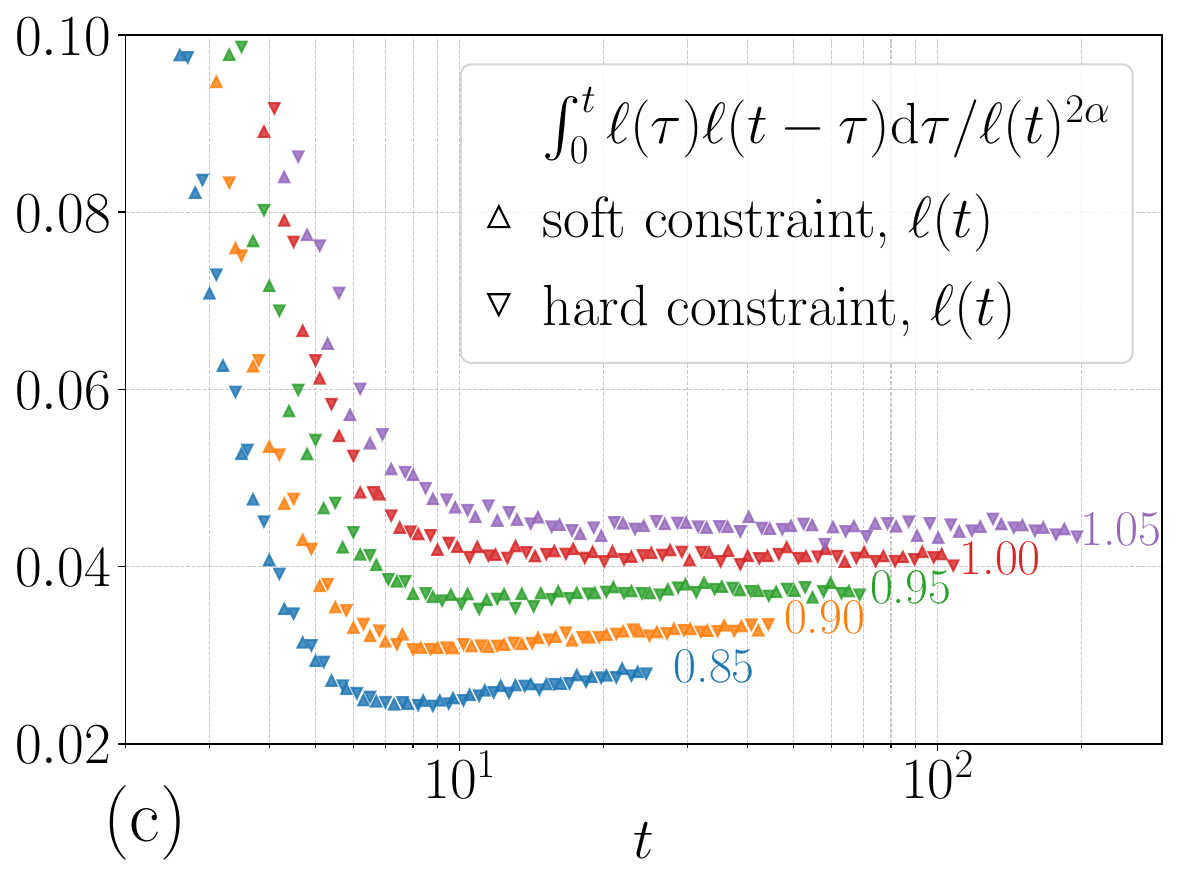}
\caption{Numerical results of $\ell(t)$ for (a) $\alpha$ on the two sides of $1.5$ (b) $\alpha  $ close to $1$ [standard log scale for the vertical axis and $|\ln(t)|^2$ scale for the horizontal axis], and (c) consistent
the check of the iterative equation for $\alpha  $ close to $1$. The dashed lines in (a) represent the functions of Fig.~\ref{fig:model_rate_phase} with adjusted prefactors. In (c), we can read that $K$ (as defined in Eq.~\ref{eq:iter_equ_2}) is roughly 24.2 when $\alpha=1$. The dashed line in (b) is Eq.~\ref{eq:criticalell} with an adjusted value for $C$, but with $c_1$ computed using $K=24.2$.}
\label{fig:ell(t)}
\end{figure*}

In the regime of a superlinear light cone ($\frac d 2 <\alpha < d +\frac{1}{2}$),  we have $\ell(t ) \gg \ell(\tau) + \ell( t - \tau) $  for any $\tau\in[0,t]$ not too close to either $0
$ or $t$. This implies that the linear size of $B_{\rm f/b}$ is much smaller than their separation. Under this approximation, the integral kernel can be simplified as $G(\|\vec{x} - \vec{y} \vec\|)  \approx \frac{1}{\ell(t)^{2\alpha}}$. Integrating over $x$ and $y$, Eq.~\ref{eq:iter_equ} reduces to an iterative integral equation of $\ell$:
\begin{equation}
\label{eq:iter_equ_2}
    \frac{1}{\ell(t)^{2\alpha}}\int_0^t  \ell^d( \tau ) \ell^d( t - \tau) \, {\rm d} \tau  \equiv \frac{1}{K} \sim 1
\end{equation}
where $K$ is a constant of order $1$. The integral can be further approximated by its value at the saddle point, which is at $\tau = \frac{t}{2}$ by symmetry. At leading order, this leads to an iterative equation $t \ell^{2d}( \frac{t}{2}) / \ell(t)^{2\alpha} \sim  \text{const.}$ \cite{hallatschek_acceleration_2014}. This simplified equation is sufficient to recover the power law and stretched exponential function in Fig.~\ref{fig:model_rate_phase}, but not the correct prefactors.  The higher order corrections and prefactors require at least Eq.~\ref{eq:iter_equ_2} for $\alpha\le d$ or at least Eq.~\ref{eq:iter_equ} for $d < \alpha < d + \frac{1}{2} $. 
We will present the higher order scalings in numerical results, but defer its details to SM\cite{SM}.

{\it Numerical results}. To bridge the missing link in our roadmap (Fig.~\ref{fig:schematic}), we numerically compare the light cone scalings of two models in $d = 1$ dimension: the hard constraint model with a maximal height $N = 2$, and the soft constraint model where the {\it equilibrium height} is approximately $2$. We define  $\ell(t)$  as the position where the average height meets a threshold $\langle h(x  =\ell(t), t )\rangle = 0.99$. We also report other statistics, such as the total population $N_p(t)$ in SM\cite{SM} --- they are all consistent with the $\ell(t)$ results.

We focus on the critical points of dimension $d = 1$ in Fig.~\ref{fig:model_rate_phase}, which separate different phases and exhibit crossover scalings. 

The light cone $\ell(t)$ increases linearly when $\alpha > 1.5$.
The system makes transition to an asymptotic power-law light cone $t^{\frac{1}{2\alpha - 2}}$ when $\alpha < 1.5$.  At the critical point $\alpha = 1.5$, the light cone scaling becomes $t \ln t$.  We demonstrate these scaling functions in Fig.~\ref{fig:ell(t)}(a); data for $\alpha = 1.3$  converges to a power-law scaling; data for $\alpha = 1.6$ and $2.0$ show slower growth consistent with a linear scaling.

There is a substantial challenge to determine the scalings near $\alpha = 1$, as the light cone follows a power law with a diverging exponent $\frac{1}{2\alpha - 2}$ on one side and a stretched exponential on the other side. 
Even with the large system size we can achieve, it remains challenging to directly confirm the asymptotic stretched exponential light cone. Instead, in Fig.~\ref{fig:ell(t)}(c), we examine whether the iterative equation Eq.~\ref{eq:iter_equ_2} has stabilized to a constant value at late times. For $\alpha \ge 0.950$, the curves appear to plateau within the accessible simulation time. 

At $\alpha = 1$, the leading order scaling is $\exp(\frac{1}{4 \ln 2} (\ln t)^2)$,   which cannot overlap well with the data of $\alpha = 1.00$ in Fig.~\ref{fig:ell(t)}(b) even at late times. We then compute the higher order saddle approximation of  Eq.~\ref{eq:iter_equ_2} (in SM\cite{SM}) and obtain the scaling up to constant order 
\begin{equation}
    \ln \ell(t) = \frac{ 1}{4 \ln 2}( \ln t - \frac{1}{2}\ln \ln t + c_1 )^2 +c_2 \ln \ln t + C +\cdots 
\label{eq:criticalell}
\end{equation}
where the constant $c_1 = \frac{1}{2} +\frac{1}{2} \ln \ln 2 + \ln K \sqrt{ \pi } $ ($K$ is defined in Eq.~\ref{eq:iter_equ_2}) and $c_2 = \frac{3}{8 \ln 2 } - \frac{1}{4} $.  In Fig.~\ref{fig:ell(t)}(b), we observe an agreement with the numerical data at late times when corrections up to $\mathcal{O}(\ln\ln t)$ are included. When plotting the prediction, we used the value of $K$ obtained from the plateau in Fig.~\ref{fig:ell(t)}(c), so that there was no fitting of unknown constants in Fig.~\ref{fig:ell(t)}(b) except for a global vertical shift in logarithmic scale.

We further perform a systematic expansion of  $\varphi = \log_2 \ell(t)$ around $\alpha = 1$, using $\delta = 2\alpha - 2$ as a small parameter. The general procedure is to approximate the integral in Eq.~\ref{eq:iter_equ_2}  at the saddle point beyond Gaussian while keeping $y = \frac\delta2\log_2t=( \alpha  - 1) \log_2 t$  fixed. This gives us an iterative equation of $\varphi$, which can be solved order by order. We derive the leading order $ \log_2\ell(t)\simeq\frac2{\delta^2} ( e^{-y } + y - 1 )$ as in \cite{hallatschek_acceleration_2014}, and produce two more orders up to vanishing terms. We compare them with the numerical data in SM\cite{SM} .

Finally, the most compelling evidence for the equivalence of the two models is that the curves for the soft and hard constraint models overlap perfectly in Fig.~\ref{fig:ell(t)}, even at relatively early times. This demonstrates that the two models are not only asymptotically identical but also match at short times for $\ell(t)$.

{\it Conclusion.} We use duality to unify the two perspectives of the operator spreading in the quantum chaotic systems. Through the duality, the birth-at-distance processes in population dynamics translate into reaction and superdiffusion terms in an FKPP equation.  While duality for a hard constraint of population limit cannot be implemented exactly, a coalescence process introducing a soft constraint of equilibrium population leads through duality to a $\frac{1}{\sqrt{N}}$ noise term in the FKPP equation. Using extensive numerical data in a 1D simulation, we demonstrate that soft and hard constraints produce identical light cone scalings, thus filling the final gap in their equivalence. 

The resulting long-range FKPP equation has a L\'evy index $\mu = 2\alpha - 2$. This contrasts the previous work\cite{zhou_hydrodynamic_2023} in which a cut-off approximation on top of the mean-field deterministic FKPP equation gives $\mu = 2\alpha - 1$. This suggests that the discreteness and the fluctuations of the height variables alter the mean field light cones differently. It will be addressed in future work. 

The power-law interaction we study in this work has broad applications in quantum simulation platforms. Current devices however are not ideal and inevitably affected by noise and decoherence. Such non-unitary effects have been observed in past experiments, such as the NMR measurement of multiple quantum coherence\cite{alvarez_localization_2011}, and they are also present in the more recent OTOC measurements in superconducting circuits\cite{mi_information_2021}. In the simplest approximation, decoherence introduces a dissipative term that drives the system toward an absorbing state of all-identity operators. This effect can be modeled using dissipative terms\cite{jacoby_spectral_2024} in the FKPP equation. It would be interesting to see if the tool of duality can be extended to the non-unitary case.

We note that the error generation and correction process in quantum devices can also be modeled as a branching long-range coalescence process\cite{odea_absorbing_2024}. Interesting error-correcting phases have been found in such models. The duality relation in this work can provide insights to the coarse-grained stochastic PDE descriptions of these phases beyond a mean-field approximation. 

\acknowledgements TZ would like to acknowledge Xiao Chen, Shenglong Xu,  Andrew Y. Guo and Brian Swingle for collaboration on related projects. TZ and XQ acknowledge Advanced Research Computing at Virginia Tech for providing computational resources and technical support that have contributed to the results reported within this paper.

\appendix

\onecolumngrid

\section{Other statistics about the light cones}

In the main text, we characterized the typical size of the light cone by $\ell(t)$, defined as the position where the average height satisfies  $\langle h( x = \ell(t) , t\rangle = 0.99$. In both the soft and hard constraint models, there are statistical observables associated with the light cone, which are not computed through the average height  $\langle h( x , t ), t\rangle$, but can be obtained from individual samples and then averaged. 

We report the results for three such observables: 
\begin{enumerate}
    \item $N_p(t) = \langle \sum_{x=-\infty}^{\infty} h(x,t) \rangle $: the total population. 
    \item $N_s(t) = \langle \sum_{x =-\infty}^{\infty}  \theta( h(x,t) ) \rangle $: the total number of occupied sites, where $\theta$  equals $1$ for positive $h$ and $0$ otherwise. 
    \item $M(t) = \Big\langle \max\Big\{ |x| \Big|   h(x, t) > 0\Big\} \Big\rangle_{\rm typical}$:  the distance to the furthest occupied site. 
\end{enumerate}

\begin{figure*}[h]
\centering
 \includegraphics[width=0.325\textwidth]{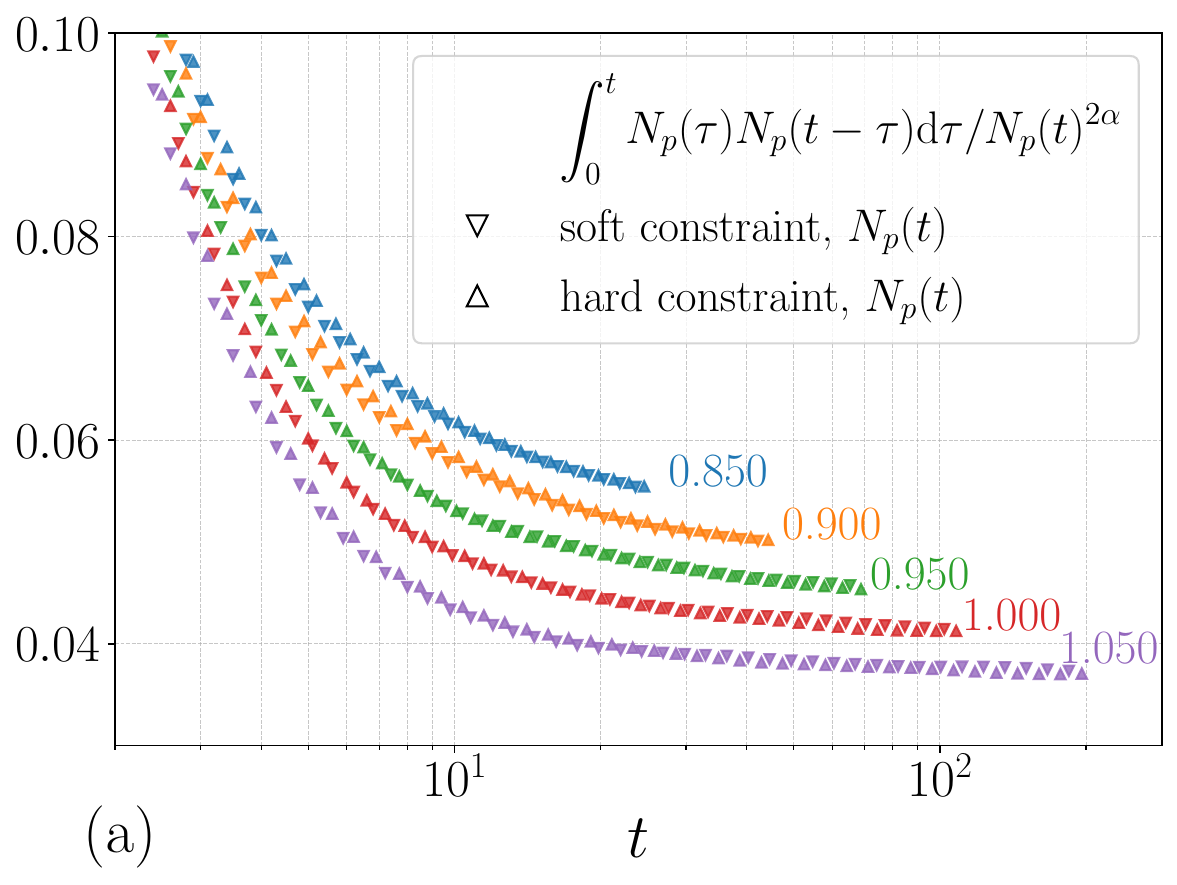}
 \includegraphics[width=0.325\textwidth]{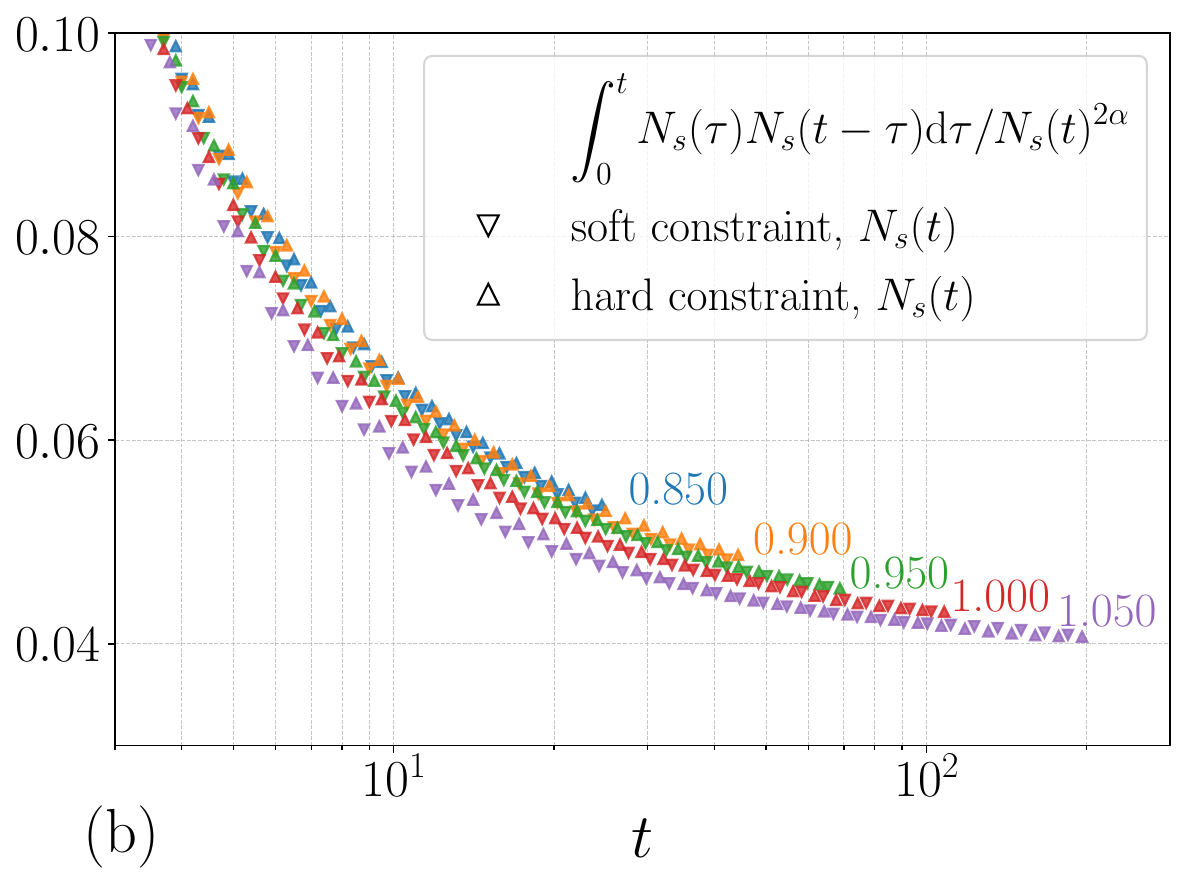}
  \includegraphics[width=0.321\textwidth]{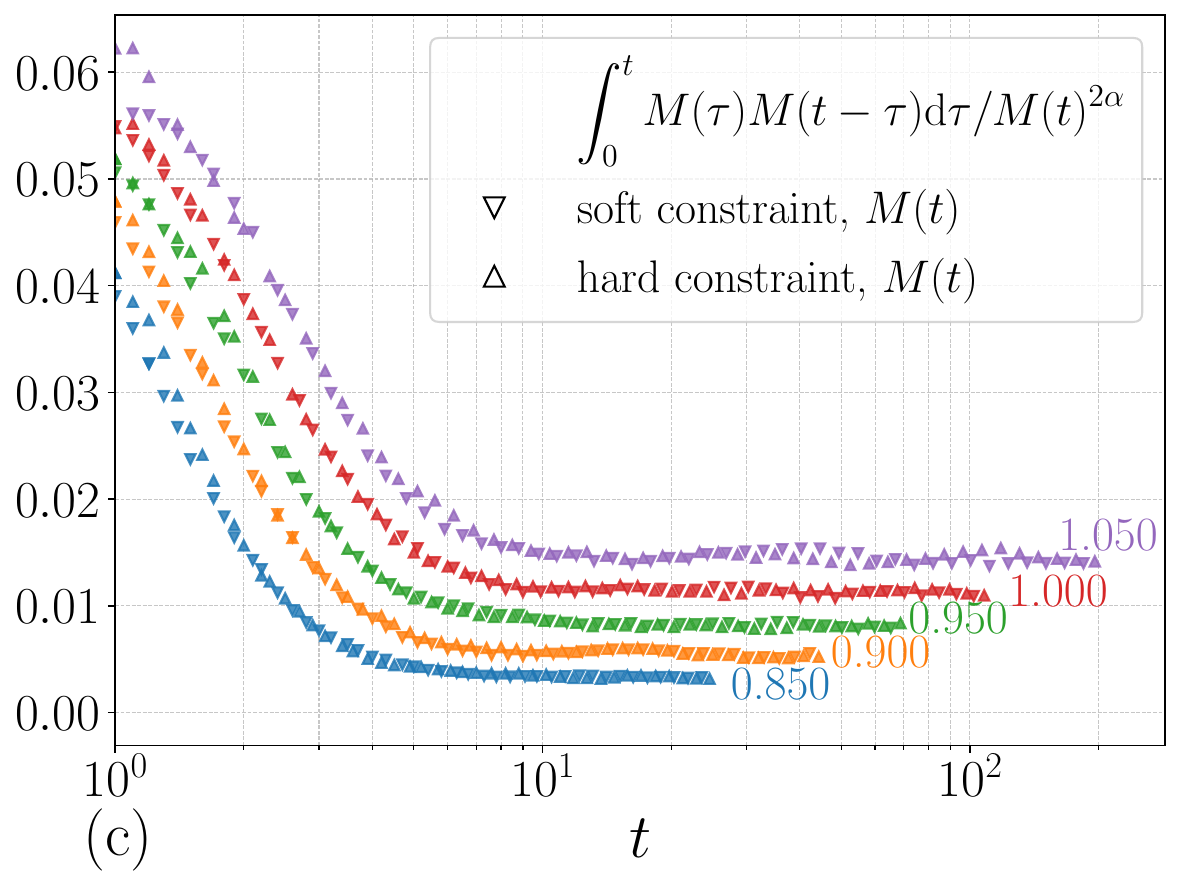}
\caption{Consistency check of the iterative integral equation against (a) $N_p(t)$,  (b) $N_s(t)$ and  (c) $M(t)$. }
\label{fig:Np_Ns_M}
\end{figure*}
The averages in $N_p(t)$ and $N_s(t)$ are taken using the standard sample mean. Since $M(t)$ involves extreme statistics, we perform a logarithmic average to capture its typical behavior. To avoid singularity at $M = 0$, we add one inside the logarithm. The typical average is defined as
\begin{equation}
    \langle \cdots \rangle_{\rm typical} 
     = \exp\left\{ \frac{1}{N_{\rm samples}}\sum_{i=1}^{N_{\rm samples} } \ln (M_i(t) +1 )  \right\} -1.
\end{equation}

We verify the iterative integral equation \eqref{eq:iter_equ_2} for $N_s(t)$, $N_p(t)$ and $M(t)$ and present the results in Fig.~\ref{fig:Np_Ns_M}. Physically, both $N_s(t)$ and $N_p(t)$ take contributions from the tails of $h(x,t)$. Their integral equations converge more slowly than $\ell(t)$. In contrast, the log-averaged typical value $M(t)$ converges much faster. We also observe that the data for the soft and hard constraint models do not exactly overlap, suggesting that microscopic mechanisms enforcing the local population limits can affect the non-universal constants of the curves. 

\section{Algorithm for simulating the height dynamics}

In this section, we describe our algorithm to simulate the height dynamics. There are popular algorithms in molecular dynamics such as the Gillespie algorithm\cite{gillespie_general_1976,gillespie_exact_1977} and tau-leaping\cite{cao_efficient_2006,gillespie_approximate_2001}. The Gillespie algorithm simulates the stochastic process exactly: it lists all the events that can occur with non-zero rates, uses the rates to compute a waiting time, and then randomly picks one of the events according to their rates to occur, updates the system, and then iterates for the next step. The tau-leaping algorithm is more efficient because it takes a larger time step by allowing multiple events to happen simultaneously, but it is an approximation. We use the exact Gillespie algorithm to check the duality. 

Below we explain our implementation of the Gillespie algorithm.  Its procedures are outlined in Alg.~\ref{alg:gill}. We record the height at discrete time steps and compute the averages at the end.  

\begin{algorithm}
\KwData{Initial state of the height: {\pcr h\_current}}
Maximal simulation time {\pcr t\_max}\\
Rates of height changing events: {\pcr rates[j]} of event {\pcr j}\\
\KwResult{ {\pcr h\_current} at each step }
\While{\pcr t < t\_max}{
  \Indp compute waiting time: {\pcr t\_waiting} $\gets$ exponential random number of average  {1 / \pcr sum(rates)}\;
  randomly select event {\pcr j}  with probability {\pcr p\_j = rates[j] / sum(rates)}\;
  update {\pcr h\_current} according to event {\pcr j}\;
  update rates according to event {\pcr j} \;
  update time: {\pcr t $\gets$ t + t\_waiting} \;
  }
\caption{Gillespie Algorithm}\label{alg:gill}
\end{algorithm}

We establish a specialized data structure to optimize the time complexity of each step in the Gillespie algorithm and eliminate the finite size effect in space. First, the birth-at-distance process has a rate for any pair of sites in which one of them has nonzero height and it is not possible to store all those events. Second, there are order $\ell(t)$ sites that can initiate a height change event at time $t$. The number of sites $\ell(t)$ can be as large as $10^9$ in order to probe the scalings for $\alpha$ close to $1$, and so we do not want the program to loop over all the sites at each timestep; our implementation allows to compute the total rate and to pick one event in constant time with $\ell(t)$ by taking advantage of the fact that, by spatial translational invariance, both the total rates for a site to initiate growth ($h\sum_r G(r)$) and to coalesce ($\frac{1}{2}\beta h(h-1)$) only depend on the local height of the site $h$. For the hard constraint model, the growth rate is also proportional to $1-\frac{h_j}{N}$  for site $j$ to increase height, but this factor can be implemented by a rejection method, see below. 

We need a data structure that allows us to perform the following operations in constant time in $\ell(t)$: 
\begin{itemize}
\item  look up the number $N_h$ of sites with a height equal to a given $h=1,2,3,\ldots$,
\item choose at random one of the $N_h$ sites with a given height $h$,
\item update the heights at any position
\item update the system in a way that all the extra information is kept consistent. 
\end{itemize}
Then the rate of the event ``one of the sites with height equal to $h$ gives birth'' is simply $\lambda N_h h$, and the rate of the event ``one of the site with height equal to $h$ coalesces'' is $\frac{1}{2}\beta N_h h (h-1)$. We can list all these events with their rates for all values of $h$ ($h$ never becomes large) and pick one at random. If the chosen event is, say, a birth from one of the sites with height equal to $h=3$, it only remains to pick at random one of the sites with height $h=3$, pick at random a jump distance, and update the system. If the chosen event is, say, a coalescence from one of the sites with height equal to $h=5$, it only remains to pick at random one of the sites with $h=5$ and update the system. In both cases, when we talk about ``updating the system'', this includes updating all the data structures that we use to make the four operations listed above easy. The time complexity of each event-picking process is thus independent of $\ell(t)$, but upper bounded by the maximal value of height, which remains small at our accessible time. In practice, we set the maximal allowable height to be $1023$ and this value has never been reached in our simulation. 

We now describe the data structure we used to implement this algorithm. (In practice, we have two independent programs with slightly different data structures allowing these operations. We only present one implementation.)
\begin{itemize}
\item The current state of height. We use the array {\pcr h\_current} to store the height for site $1 \le x \le L$. (We take the initial state to be $h(\frac{L}{2}, t=0 ) = 1$ and $0$ elsewhere to avoid boundary effects. However, in the numerical examples shown below in the figures, we place the nonzero-height cluster near site 1 for the clarity of the presentation.) We use a dictionary (hash table) to store sites with heights outside the above range. The array ensures fast random access of the height and data recording for computing the average height profile. The dictionary records the relatively rare long-distance jumps with the same $\mathcal{O}(1)$ access performance and reasonable overhead. Most of the occupied sites (non-zero heights) should be within the range of $1 \le x \le L$, where $L \sim 10^8$. 
\begin{center}
\includegraphics[width=0.8\columnwidth]{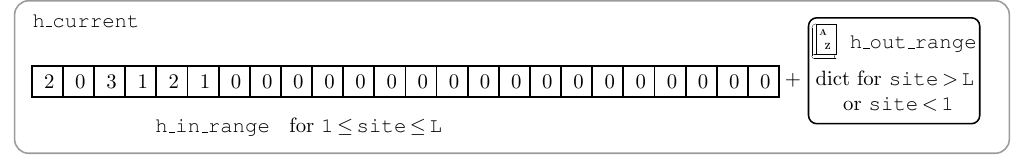}
\end{center}
\item  The structure {\pcr site\_of\_rank} is a list of the all the indices of non-empty sites in {\pcr h\_current} sorted in decreasing order of $h$. If, for instance, the largest height is $h=3$, then it stores all the indices of sites with $3$ particles, then all the indices of sites with 2 particles, and finally all the indices of sites with 1 particle.
\item  The structure {\pcr rank\_of\_site} is the reverse map of {\pcr site\_of\_rank} -- it stores the rank (in the {\pcr site\_of\_rank} list) of a non-empty site at position $x$. In other words, if {\pcr h\_current[$x$]}$\ge 1$, then {\pcr site\_of\_rank[rank\_of\_site[$x$]]==$x$}.  The rank of a height zero is invalid and stored as $0$. {\pcr rank\_of\_site} is implemented as an array for in-range sites and dictionary for out-of-range sites, for the same reason as {\pcr h\_current}. 
\begin{center}
\includegraphics[width=0.8\columnwidth]{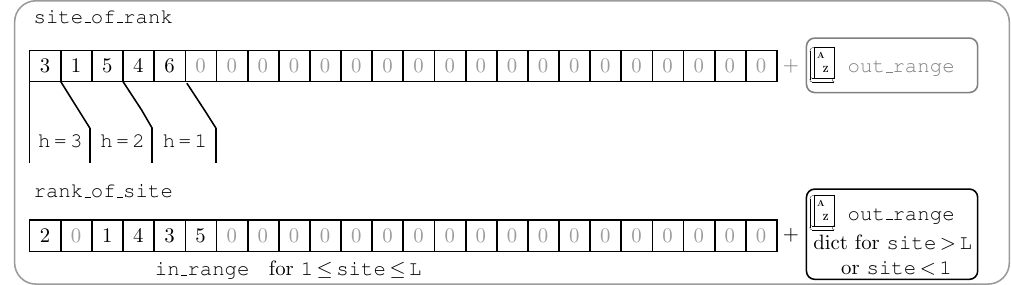}
\end{center}
\item We use {\pcr h\_count} to record the number of sites with a given non-zero heights $h$, and {\pcr h\_inc\_rate}, {\pcr h\_dec\_rate} to record the total rates of height increasing and decreasing. These vectors are updated every time when a height change occurs. The total number of events can reach $10^{10}$ in a simulation. The rates are integer multiples of the total birth rate $\lambda  = G(0) + 2\sum_{r >0} G(r)$ and of the death rate $\beta$; to maintain accuracy and avoid rounding errors, we only store those integers and multiply them by $\lambda$ or $\beta$ when needed. 
\begin{center}
\includegraphics[width=0.8\columnwidth]{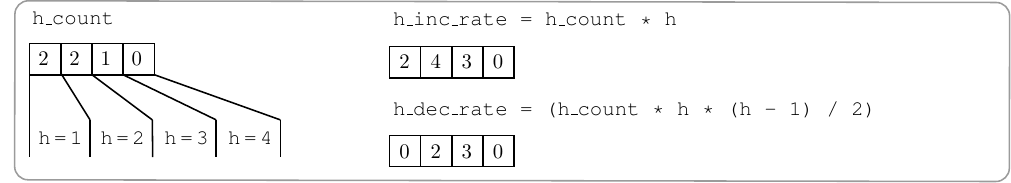}
\end{center}
\end{itemize}

A single time step of the algorithm is implemented as follows: 
\begin{enumerate}
\item Update the time of the system after the current step. To do this, compute the total rate, equal to
\begin{equation}
    R=\sum_h \big(\text{\pcr h\_inc\_rate}[h]\times\lambda +\text{\pcr  h\_dec\_rate}[h]\times\beta\big)
\end{equation} 
and increase the time  by a random exponential number of average $1/R$.
\item Randomly select an event with probability proportional to the rates
\begin{enumerate}
\item[a)] We randomly select whether we do a birth or a coalescence, and the height $h$ of the site where the event starts, according to the total rates for that height from {\pcr h\_inc\_rate} and {\pcr h\_dec\_rate}. In this step, we do not yet select on which site the event starts. The total rate of all the possible births starting from any site with height $h$ is {\pcr h\_count}$[h]\times h \times \lambda =\text{\pcr h\_inc\_rate}[h]\times\lambda $, and the total rate of all the coalescence happening on any site with height $h$ is {\pcr h\_count}$[h](${\pcr h\_count}$[h]-1)\times\beta=\text{\pcr  h\_dec\_rate}[h]\times\beta$. In the example below, we have selected the event where a (yet unknown) site of height $1$ is to increase the height of another site. 
\begin{center}
\includegraphics[width=0.8\columnwidth]{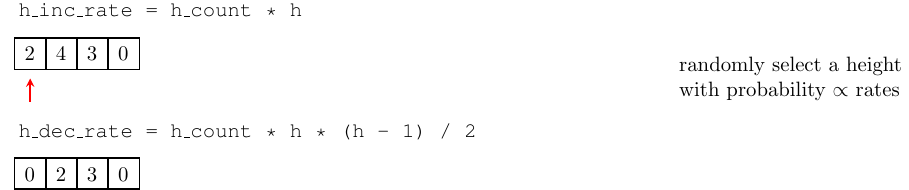}
\end{center}
\item[b)] Then, we uniformly pick one of the sites with the selected height $h$ to be the site where the event starts. To do this, we use {\pcr h\_count} to determine the ranks of the relevant sites in the {\pcr site\_of\_rank} array: there are $\sum_{h'\ge h+ 1}$ {\pcr h\_count}$[h']$ sites before them (in the example, there are three: two with $h=2$ and one with $h=3$), and we know how many sites with the selected height there are (in the example there are two sites with $h=1$), and so we can obtain the range of ranks of the sites with the selected height. (In the example, the $h=1$ sites have ranks 4 to 5.) We choose one of them uniformly. (Say we choose the one at rank 5.)
\begin{center}
\includegraphics[width=0.8\columnwidth]{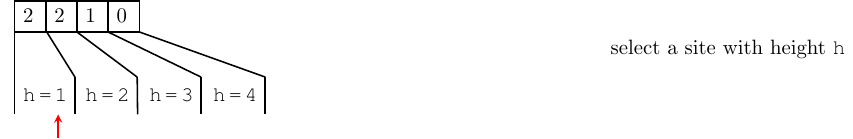}
\end{center}

\item[c)] We read in {\pcr               site\_of\_rank} the spatial position of the selected site from its rank. In the example, the particle with rank 5 is at position~6 in the system; this is where the new event starts from. In the case of a coalescence, we would decrease the height of that site by one unit. In the case of a birth, we would increase by one unit the height of a site at some random distance $r$ from there. In the example we do a birth at distance; we pick randomly the value of $r$ and the direction, let us say we obtain $r=1$ to the left, and we update the corresponding site; here the site at position~5.
\end{enumerate}
\item We know which site to increase or decrease; in the example, we increase the site at position~5. We must update the height status in all the recording structures
\begin{enumerate}
\item[a)] We update the height in {\pcr h\_current}. 
\begin{center}
\includegraphics[width=0.8\columnwidth]{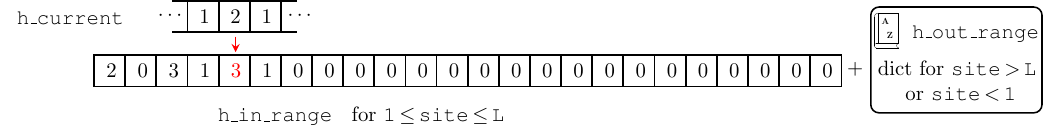}
\end{center}
\item[b)] We update the rank of the modified site. We know its old rank from the {\pcr rank\_of\_site} array (rank~3 in the example), and we know from {\pcr h\_count} the range of ranks for all the sites having the old height of the modified site. (In the example, the old height is 2, and all the sites with height 2 have ranks ranging from 2 to 3). We need to put the modified site at the beginning of that range in the case of an increase (as in the example) or at the end in the case of a decrease. In the example, the rank of the modified height is~3, we need it to be 2, so we exchange the sites at ranks 2 and 3. We must do the change in both the {\pcr rank\_of\_site} and {\pcr site\_of\_rank} arrays so that they remain consistent.
\begin{center}
\includegraphics[width=0.8\columnwidth]{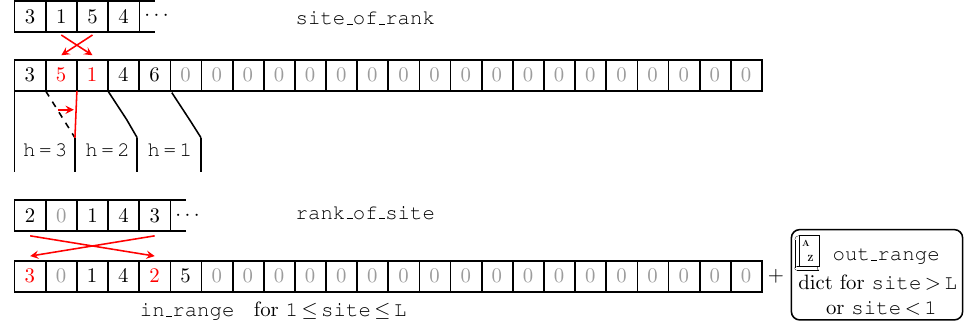}
\end{center}
\item[c)] Finally we update the number of the sites with the height counts, and the corresponding rates. After updating {\pcr h\_count} as in the example, the site with rank 2 (which corresponds to the modified height) is now in the range of ranks corresponding to a height 3.
\begin{center}
\includegraphics[width=0.8\columnwidth]{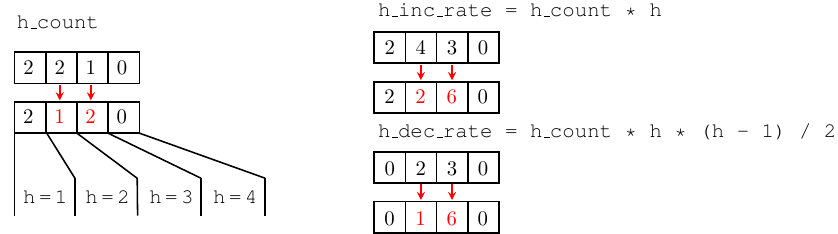}
\end{center}
\end{enumerate}
\end{enumerate}

In the simulation, we restrict the maximal height to be $h_{\rm max} = 1023$. In practice, no instances have reached this height, therefore it does not impose a constraint in the simulation. The data structure and algorithm above achieved $\mathcal{O}(h_{\rm max})$ time complexity to select a event, and $\mathcal{O}(1)$ time to update the book-keeping array. 

We have just described the algorithm for the soft constraint model. For the hard constraint model, we need to make two small changes: first, we set $\beta$ to 0 so that there are no coalescence. Then, when a birth at distance is scheduled as described above, only execute the change with probability $1-\text{\pcr h\_current}[j]/N$, where $\text{\pcr h\_current}[j]$ is the height of the site where the increase is to take place and $N$ is the upper limit of the local population (the hard constraint). If the test fails, the change is rejected, and nothing is updated except the time of the system (step 1).

\section{Adjusting the equilibrium height}

The transitions in the hard constraint model only increase the height. Therefore the height of any site is a non-decreasing function over time, and eventually, it will saturate at the maximal height $N$. On the other hand, height can both increase and decrease in the soft constraint model. To have a fair comparison of their light cones, it is favorable to adjust the equilibrium height of the soft constraint model to be equal to $N$. We computed the corresponding coalescence rate below. 

Consider the soft constraint model in 1D, and $h_i(t)$ is the height of site $i$ at time $t$. The transition rates of the height change at a sample site $0$ is given by 
\begin{equation}
h_0(t+\diffd t) = \begin{cases}
    h_0(t) +1 &\text{proba. $\sum_i G(|i|) h_i(t)\,\diffd t$},\\
    h_0(t) -1 &\text{proba. $\beta \frac{h_0(t)(h_0(t)-1)}2\,\diffd t$},\\
    h_0(t)    &\text{proba. $1-\Big(\sum_i G(|i|) h_i(t)+\beta \frac{h_0(t)(h_0(t)-1)}2\Big)\diffd t.$}
\end{cases}
\end{equation}
Here we use $\beta$ to denote the rate at which two particles on the same site coalesce, and $G( | i -j | )$ to denote the rate at which a particle at site $i$ gives birth at site $j$. The notation $G(|i-j|)$ implies translation invariance and reflection symmetry of the function, so that the total birth rate can be written as
\begin{equation}
\lambda = \sum_{i=-\infty}^{\infty} G(|i|).
\end{equation}

The mean field solution would be
\begin{equation}
    \frac{\diffd \langle h_0\rangle }{\diffd t} = \sum_i G(|i|)\langle h_i\rangle -\beta \frac{\langle h_0\rangle(\langle h_0\rangle-1)}{2} .
\end{equation}
At time $t \rightarrow \infty$, $\langle  h_i(t) \rangle \rightarrow N $, and $\frac{\diffd \langle h_0 \rangle }{dt} \rightarrow 0$,  which suggests $\beta = \frac{2\lambda}{N-1} $. However, for $N = 2$  and $\beta = 2 \lambda$, we observe numerically that $\langle h_i (t) \rangle $ converges to about $1.7$, which is consistent with our prediction below. The mean field coalescence rate is too strong. 

To take into account the fluctuation, define the generating function 
\begin{equation}
\Phi(q,t)=\langle q^{h_0(t)}\rangle \equiv \sum_{n=0}^{\infty} q^{n} \mathbb{P}[h_0(t) = n ].
\end{equation}

Its change in an infinitesimal time is
\begin{equation}
\begin{aligned}
\Phi(q,t+\diffd t) &= \Big\langle q^{h_0(t)}\Big((q-1)\sum_i G(|i|)h_i(t)\diffd t+ ( q^{-1}-1) \beta \frac{h_0(t)(h_0(t)-1)}2\diffd t+1\Big)\Big\rangle
\\&=\Phi(q,t)+ \Phi(q,t)(q-1)\sum_i G(|i|)\langle h_i(t)\rangle\diffd t + (q^{-1}-1)\beta\frac12 q^2 \frac{\partial^2 \Phi}{\partial q^2}\diffd t
\end{aligned}
\end{equation}
where we have neglected the spatial correlation $\langle q^{h_0(t) } h_i(t)  \rangle =\langle q^{h_0(t) } \rangle \langle  h_i(t)  \rangle $, a valid assumption at late times. The equation for $\Phi$ is 
\begin{equation}
\frac{\partial \Phi}{\partial t} = (q-1) \Big[\Phi\sum_iG(|i|)\langle h_i(t)\rangle -\frac{\beta q}2\frac{\partial^2 \Phi}{\partial q^2}\Big]
\end{equation}
When $t \rightarrow \infty$, $\Phi(q,t)\to \Phi_\text{eq}(q)$ and $\langle h_i(t)\rangle\to \langle h\rangle_\text{eq}$ , independent of position. The equilibrium distribution of particles on site 0 is therefore characterized by $\Phi_\text{eq}$ , which is the solution to
\begin{equation}
\lambda\langle h\rangle_\text{eq} \Phi_\text{eq} -\frac{\beta q}2\frac{\partial^2 \Phi_\text{eq}}{\partial q^2}=0
\end{equation}
The general solution is
\begin{equation}
\Phi_\text{eq}(q) = A \sqrt q I_1\Big(\sqrt{\tfrac{8\lambda\langle h\rangle_\text{eq}q}\beta}\Big)+B \sqrt q K_1\Big(\sqrt{\tfrac{8\lambda\langle h\rangle_\text{eq}q}\beta}\Big)
\end{equation}
where $I_1$ and $K_1$ are Bessel functions. Even in the soft constraint model, coalescence can not bring it down to 0 once the height is nonzero. Therefore the probability of having $h_0(t) = 0$ at $t \rightarrow \infty$  is zero. It is legitimate to take the limit $\lim_{q \rightarrow 0} \Phi_\text{eq}(q)=0$. This condition eliminates $B$. Furthermore, the total probability is normalized to $1$, which gives  $\Phi_\text{eq}(1)=1$ and 
\begin{equation}
\Phi_\text{eq}(q)= \frac{I_1\Big(\sqrt{\tfrac{8\lambda\langle h\rangle_\text{eq}q}\beta}\Big)}{I_1\Big(\sqrt{\tfrac{8\lambda\langle h\rangle_\text{eq}}\beta}\Big)}\sqrt q.
\end{equation}
Then, since at any time $\langle h_0(t)\rangle =\frac{\diffd}{\diffd q}\Phi(q,t)\big|_{q=1}$, one has 
\begin{equation}
\label{eq:rate_to_set_h_eq}
\langle h\rangle_\text{eq} = \Phi_\text{eq}'(1)=\frac12\left(1+\sqrt{\tfrac{8\lambda\langle h\rangle_\text{eq}}\beta}\frac{I_1'\Big(\sqrt{\tfrac{8\lambda\langle h\rangle_\text{eq}}\beta}\Big)}{I_1\Big(\sqrt{\tfrac{8\lambda\langle h\rangle_\text{eq}}\beta}\Big)}\right)
\end{equation}
at equilibrium. 

Eq.~\ref{eq:rate_to_set_h_eq} is the relation to adjust $\beta$ and $\lambda$ to set $\langle h \rangle_{\rm eq} = N$.  When we set $\langle h \rangle_{\rm eq} = 2$
\begin{equation}
3= \sqrt{\tfrac{16\lambda}\beta}\frac{I_1'\Big(\sqrt{\tfrac{16\lambda}\beta}\Big)}{I_1\Big(\sqrt{\tfrac{16\lambda}\beta}\Big)}.
\end{equation}
Numerically, this gives $\sqrt{\tfrac{16\lambda}\beta}=3.32585$, or $\beta=1.44649\lambda$. In the numerical simulation of the soft constraint, we have $\langle  h \rangle_{\rm eq}$ to be equal to $2$ with less than $1\%$ error. 

On the other hand, if we take the mean field result $\beta=2\lambda$ (for $N = 2$), then
\begin{equation}
\langle h\rangle_\text{eq}=\frac12\left(1+2\sqrt{\langle h\rangle_\text{eq}}
\frac{I_1'\Big(2\sqrt{\langle h\rangle_\text{eq}}\Big)} {I_1\Big(2\sqrt{\langle h\rangle_\text{eq}}\Big)}\right).
\end{equation}
Numerically this gives $\langle h\rangle_\text{eq}=1.66978.$

\section{Duality}

In this section, we explain how one can construct a dual continuous process to certain Markovian particle systems. Most of the ideas in this appendix were already exposed in
\cite{doering_interacting_2003}. However, \cite{doering_interacting_2003} did not consider the case of a birth at a distance which we need in this work.

To simplify notations, we shall explain the ideas behind duality on a lattice made of only two sites. The generalization to a larger lattice is immediate.

\subsection{Master equation for a particle system}

We consider a particle system living on two sites, labeled 1 and 2. At any time $t$, the state of the system is characterized by the heights (or numbers of  particles) $h_1(t)$ and $h_2(t)$ on each of the two sites. To shorten notations, we write possible states, or configurations, of the system as two-component vectors such as $\vec m$ or $\vec n$. In particular, we write $\vec h(t)=\begin{spmatrix}h_0(t)\\h_1(t)\end{spmatrix}$ for the state of the system at time $t$.

Let $P(\vec n,t)$ be the probability that the system at time $t$ is in the configuration $\vec n$, \textit{i.e.}:
\begin{equation}
P( \vec{n}, t ) = \proba( \vec{h}(t) = \vec n ) .
\end{equation}
As we assume the system to be Markovian, the distribution of the state at time $t+ \diffd t$ only depends on the state at time $t$, and one can write a master equation of the form
\begin{equation}\label{eq:masterP}
  \partial_t P( \vec{n}, t ) = \sum_{\vec{m}} M_P(\vec{n}, \vec{m}) P( \vec{m}, t ),
\end{equation}
where the ``matrix'' $M_P(\vec{n}, \vec{m})$ encodes all the possible transitions in the system. To understand this, it is better to write \autoref{eq:masterP} in the following equivalent way:
\begin{equation}
P(\vec n, t+\diffd t) = \big(1+M_P(\vec n,\vec n)\diffd t\big)P(\vec n,t) + \sum_{\vec m\ne\vec n} M_P(\vec n,\vec m)P(\vec m,t)\diffd t,
\end{equation}
which one reads as: ``the probability to be in state $\vec n$ at time $t+\diffd t$ is the probability $P(\vec n,t)$ to be already in state $\vec n$ at time $t$ and to remain there during $\diffd t$ (with probability $1+M_P(\vec n,\vec n)\diffd t$), plus the sum over all the other states $\vec m$ of the probability $P(\vec m,t)$ to be in state $\vec m$ at $t$ and to transition from $\vec m$ to $\vec n$ during $\diffd t$ (with probability $M_P(\vec n,\vec m)\diffd t$.''
Therefore, $M_P(\vec n,\vec m)$ with $\vec n\ne\vec m$ is the rate at which the system may transition from $\vec m$ to $\vec n$, and $M_P(\vec m,\vec m)$ is minus the rate at which the system leaves state $\vec m$. This implies that, for all $\vec m$,
\begin{equation}
\sum_{ \vec{n}}M_P(\vec n,\vec m) = 
\Big(\sum_{ \vec{n}\ne \vec{m}} M_P( \vec{n}, \vec{m} )\Big) - \big(- M_P ( \vec{m}, \vec{m} )\big) = 0,
\label{eq:propMP}
\end{equation}
because both terms in the difference are two ways of writing the rate of leaving state $\vec m$.
With this relation, one can check in \autoref{eq:masterP} that probabilities remain normalized, \textit{i.e.} $\partial_t\sum_{\vec n}P(\vec n,t)=0$.

\subsection{Coupled diffusive processes}
We now consider coupled diffusive processes $Z_1(t)$ and $Z_2(t)$ (one process for each of the two lattice sites) which are real numbers evolving according to stochastic differential equations of the form
\begin{equation}
\diffd Z_i =  \mu_i(Z_1, Z_2 )\,\diffd t + {\sigma_i(Z_1,Z_2)}\,\diffd B_i.
\label{eq:sde}
\end{equation}
In this equation, \begin{itemize}
\item the functions  $ B_i(t)$ are independent Brownian motions for each site $i$; in other words, the variations $\diffd  B_i$ of $B_i$ during $\diffd t$ are independent Gaussian numbers for different values of $i$ and $t$ such that $\langle \diffd B_i\rangle=0$ and $\langle (\diffd B_i)^2\rangle=\diffd t$.

\item the quantities $\mu_i(Z_1,Z_2)$ and $\sigma_i(Z_1,Z_2)$ are respectively the drift and the diffusion coefficient of the process $Z_i(t)$. They depend upon the values $Z_1(t)$ and $Z_2(t)$ of both diffusions at time $t$.
\end{itemize}

For a vector $\vec n=\big(\begin{smallmatrix}n_1\\n_2\end{smallmatrix}\big)$ of two integers $n_1$ and $n_2$, we introduce the joint moments $Q(\vec n,t)$ of the $Z$ as
\begin{equation}
  Q( \vec{n}, t ) = \Big\langle Z_1(t)^{n_1}Z_2(t)^{n_2}\Big\rangle.
\end{equation}
We seek to write an equation for the time evolution of $Q$. To do so, we need to be careful and remember that the term $\diffd B_i$ in \autoref{eq:sde} is of order $\sqrt{\diffd t}$; in other words, we need to use Itô calculus. We start by 
\begin{equation}
Q(\vec n,t+\diffd t)  = \Big\langle 
Z_1(t+\diffd t)^{n_1}
Z_2(t+\diffd t)^{n_2}
\Big\rangle 
\label{eq:Qt+dt}
\end{equation}
where, after Taylor expansion, ignoring terms smaller than $\diffd t$,
\begin{equation}
Z_i(t+\diffd t)^{n_i} = \big[ Z_i(t) + \diffd Z_i
\big]^{n_i}
= Z_i(t)^{n_i} + n_i Z_i(t)^{n_i-1}\diffd Z_i + \frac{n_i(n_i-1)}{2}Z_i(t)^{n_i-2}(\diffd Z_i)^2
\end{equation}
According to Itô's rule, the $(\diffd   Z_i)^2
$ in the last term may be replaced by its average $\sigma_i^2\,\diffd t$ and one obtains
\begin{equation}
\label{eq:Zt_+_dt}
\begin{aligned}
Z_i(t+\diffd t)^{n_i}
&= Z_i(t)^{n_i} + \Big[ \mu_i n_i Z_i(t)^{n_i-1}+\sigma_i^2 \frac{n_i(n_i-1)}{2}Z_i(t)^{n_i-2}\Big]\diffd t +  \sigma_i n_i Z_i(t)^{n_i-1} \,\diffd B_i,
\\
&= Z_i(t)^{n_i}\bigg(1 + \Big[ \frac{\mu_i n_i}{ Z_i(t)}+\sigma_i^2 \frac{n_i(n_i-1)}{2Z_i(t)^2}\Big]\diffd t +  \frac{\sigma_i n_i}{ Z_i(t)} \,\diffd B_i\bigg).
\end{aligned}
\end{equation}
(For brevity, we write $\mu_i$ and $\sigma_i$ instead of $\mu_i(Z_1,Z_2)$ and $\sigma_i(Z_1,Z_2)$.)

We plug \autoref{eq:Zt_+_dt} in \autoref{eq:Qt+dt}; first we notice when expanding the products that all the $\diffd B$ terms will disappear when taking the expectation because we assumed that $\langle\diffd B_i\rangle=0$ and $\langle \diffd B_1\diffd B_2\rangle=0$. We therefore omit all those terms and obtain, to leading order in $\diffd t$
\begin{equation}
Q(\vec n,t+\diffd t)  = \bigg\langle Z_1(t)^{n_1}Z_2(t)^{n_2}\bigg(1+\diffd t\,\sum_{i=1,2}   \Big[\frac{\mu_i n_i}{Z_i(t)}+\frac{\sigma_i^2 n_i(n_i-1)}{2Z_i(t)^2}\Big]\bigg)\bigg\rangle,
\end{equation}
or, equivalently,
\begin{equation}
\partial_tQ(\vec n,t)  = \bigg\langle Z_1(t)^{n_1}Z_2(t)^{n_2}\sum_{i=1,2}   \Big[\frac{\mu_i n_i}{Z_i(t)}+\frac{\sigma_i^2 n_i(n_i-1)}{2Z_i(t)^2}\Big]\bigg\rangle.
\label{eq:dQ}
\end{equation}
We assume that the drifts $\mu_i(Z_1,Z_2)$ and the squared diffusion coefficients $\sigma_i(Z_1,Z_2)^2$ can be written as polynomials of $Z_1$ and $Z_2$. Then the right hand side of \autoref{eq:dQ} is the expectation of a sum of terms of the form $Z_1(t)^{m_1}Z_2(t)^{m_2}$, for several values of the vector $\vec m=\big(\begin{smallmatrix}m_1\\m_2\end{smallmatrix}\big)$. In other words, \autoref{eq:dQ} may be rewritten as
\begin{equation}\label{eq:masterQ}
\partial_t Q( \vec{n}, t ) = \sum_{\vec{m}} M_{Q} ( \vec{n}, \vec{m} ) Q( \vec{m} , t ) 
\end{equation}
for some matrix $M_Q$.
This looks very similar to \autoref{eq:masterP}, but one must remember that $Q$ is not a distribution of probability, it is not normalized, and the matrix $M_Q$ does not satisfy \autoref{eq:propMP}.

\subsection{Duality}
We now consider, on the same two-site lattice, a particle system $h_i(t)$ described by the matrix $M_P$ and coupled diffusive processes $Z_i(t)$ with moments described by the matrix $M_Q$. The processes are independent. Let $t>0$ be some arbitrary positive time, and introduce, for $0\le \tau \le t$,
\begin{equation}\label{eq:Atau}
  A(\tau) = \Big\langle Z_1(\tau)^{h_1(t-\tau)}  Z_2(\tau)^{h_2(t-\tau)} \Big\rangle_{h,Z}
= \Big\langle Q\big(\vec h(t-\tau), \tau\big)\Big\rangle_{h} = \sum _{\vec n}P(\vec n,t-\tau) Q(\vec n,\tau).
\end{equation}
(In the first expression for $A(\tau)$, the expectation is made on both the $h$ and the $Z$ processes, while in the second one, it is only made on the $h$ process.)

We say that the processes $h$ and $Z$ are dual if the quantity $A(\tau)$ is independent of $\tau$ for any choice of $t$ and any choice of the initial conditions $h_i(0)$ and $Z_i(0)$. In particular, assuming that the initial conditions are non-random, writing $A(0)=A(t)$ gives
\begin{equation}
\Big\langle Z_1(0)^{h_1(t)}  Z_2(0)^{h_2(t)} \Big\rangle_{h}
=\Big\langle Z_1(t)^{h_1(0)}  Z_2(t)^{h_2(0)} \Big\rangle_{Z}
\label{eq:du}
\end{equation}
which is like \autoref{eq:duality_0_tau} after taking $Z=1-f$.

To find the condition for duality, we compute $\partial_\tau A(\tau)$ from the last expression in \autoref{eq:Atau}, using \autoref{eq:masterP} and \autoref{eq:masterQ}:
\begin{equation}
\begin{aligned}
  \partial_\tau A(\tau) &= 
\sum_{\vec n}\Big[-\partial_t P(\vec n,t-\tau) Q(\vec n,\tau) + P(\vec n,t-\tau)\partial_t Q(\vec n,\tau)\Big],
\\&=
\sum_{\vec n}\Big[-\Big(\sum_{\vec m} M_P(\vec n,\vec m) P(\vec m,t-\tau)\Big) Q(\vec n,\tau) + P(\vec n,t-\tau)\Big( \sum_{\vec m} M_Q(\vec n,\vec m)Q(\vec m,\tau)\Big)\Big],
\\&=
 - \sum_{\vec{n}, \vec{m}} M_{P} ( \vec{n}, \vec{m} ) P( \vec{ m}, t-\tau ) Q( \vec{n}, \tau ) 
+\sum_{\vec{n}, \vec{m}} M_Q( \vec{n}, \vec{m} )P( \vec{n},  t-\tau )  Q( \vec{m}, \tau ),
\\ &=
\sum_{\vec{n}, \vec{m}} \big(M_Q( \vec{n}, \vec{m}\big ) - M_P( \vec{m}, \vec{n}) ) P(\vec{n}, t- \tau) Q( \vec{m}, \tau )  .
\end{aligned}
\end{equation}
Duality is satisfied when
\begin{equation}
M_Q( \vec{n}, \vec{m} ) = M_P( \vec{m}, \vec{n} ),
\end{equation}
\textit{i.e.}\@ when the matrices $M_Q$ and $M_P$ are the transpose of each other.

Therefore, one can try to construct the dual of a given particle systems by first writting $M_P$, then $M_Q=M_P^{\top}$, and then by identifying the corresponding $\mu$ and $\sigma$ of the diffusive processes.
We show how to do this in the next two sections, first on simple examples and then on the OTOC height evolution system.

\subsection{Simple examples}

\subsubsection{Birth on site}\label{sec:birthonsite}
Consider a particle system on a single site, where the number $h$ of particles on the site increases by one with rate $\gamma h$. (This is a Galton-Watson process in continuous time.) The non-zero elements of the matrix $M_P$ are
\begin{equation}
M_P( n+1, n ) = \gamma n,  \quad M_P( n, n ) = - \gamma n \quad\text{for all $n=1,2,3,\ldots$}
\end{equation}
(the first expression is the rate going from $n$ to $n+1$, and the second is minus the rate of leaving $n$.) Then, the non-zero elements of the $M_Q$ matrix are

\begin{equation}
M_Q( n, n+1 ) = \gamma n  \quad M_Q( n, n ) = - \gamma n \quad\text{for all $n=1,2,3,\ldots$}.
\end{equation}
Then, \autoref{eq:masterQ} becomes
\begin{equation}
\partial_t Q(n,t) =\sum_m M_Q(n,m) Q(m,t)=\gamma n Q(n+1,t)-\gamma n Q(n,t)=\bigg\langle Z(t)^n\Big[\gamma n Z(t)-\gamma n\Big]\bigg\rangle.
\end{equation}
With only one site in the system, \autoref{eq:dQ} becomes
\begin{equation}\label{eq:dQ2}
\partial_t Q(n,t) =  \bigg\langle Z(t)^{n}\Big[\frac{\mu(Z(t)) n}{Z(t)}+\frac{\sigma(Z(t))^2 n(n-   1)}{2Z(t)^2}\Big]\bigg\rangle.
\end{equation}
Therefore, by taking
\begin{equation}
\mu(Z) = \gamma (Z^2  - Z ) ,\qquad \sigma(Z)=0,
\end{equation}
the processes $h$ and $Z$ are dual. In this specific case, $Z$ is actually deterministic since $\sigma=0$, and \autoref{eq:sde} becomes $\diffd Z = \gamma(Z^2-Z)\,\diffd t$. Solving with $Z(0)=q$ then gives in \autoref{eq:du}  the generating function for $h(t)$:
\begin{equation}
\big\langle q^{h(t)}\big\rangle= \bigg(\frac{q}{e^{\gamma t}(1-q)+q}\bigg)^{h(0)}.
\end{equation}

\subsubsection{Coalescence on site}\label{sec:coalesce}
Consider a particle system on a single site, where the number $h$ of particles on the site decreases by one with rate $\beta \frac{h(h-1)}2$. The non-zero elements of the matrices $M_P$ and $M_Q$ are 
\begin{equation}\begin{aligned}
M_P( n-1, n ) &= \beta \frac{n(n-1)}{2}  &M_P( n, n ) &= - \beta \frac{n (n - 1)}{2}&\text{for all $n=2,3,4,\ldots$}\\
 M_Q( n, n-1 ) &= \beta \frac{n (n-1)}{2}  & M_Q( n, n ) &= - \beta \frac{n(n-1)}{2} &\text{for all $n=2,3,4, \ldots$},
\end{aligned}
\end{equation}
so that
\begin{equation}
\partial_t Q(n,t) =\sum_m M_Q(n,m) Q(m,t)=\beta\frac{n(n-1)}2 \Big[Q(n-1,t)- Q(n,t)\Big]=
\beta\frac{n(n-1)}2\bigg\langle  Z(t)^n\Big[\frac1{Z(t)}-1\Big]\bigg\rangle.
\end{equation}
Comparing with \autoref{eq:dQ2}, we see that we must take
\begin{equation}
\mu(Z)=0,\qquad\sigma(Z)^2= \beta ( Z - Z^2 ).
\end{equation}
The resulting stochastic differential equation is the neutral Wright-Fisher diffusion $\diffd Z = \sqrt{\beta (Z-Z^2)}\,\diffd B$ ; this is used to describe the proportion of a population carrying a given neutral mutation. Starting from a given value $Z(0)\in(0,1)$, the quantity $Z(t)$ evolves randomly until it hits either 0 (the mutation has disappeared) or 1 (the mutation has fixated), and then it remains constant. The process $h$, obviously, decreases from $h(0)$ to 1, and remains at 1. From the duality relation, one has
\begin{equation}
\langle Z(0)^{h(t)} \rangle = \langle Z(t)^{h(0)}\rangle.
\end{equation}
Sending $t\to\infty$, this leads to the classic result that the probability of fixation is $Z(0)$:
\begin{equation}
Z(0) = \P[Z(\infty)=1].
\end{equation}

\subsubsection{Birth and coalescence on a site}
Consider a particle system on a single site where both transitions of the two previous examples are present: the population can increase by 1 with rate $\gamma h$ and decrease by one with rate $\beta\frac{h(h-1)}2$.
We don't have to do any more work; the matrix $M_P$ of this process is simply the sum of the matrices $M_P$ of the two previous cases; this will also be true for the matrix $M_Q$, and the parameters $\mu$ and $\sigma^2$. Therefore, this process is dual to a diffusion with parameters
\begin{equation}
\mu(Z)=\gamma(Z^2-Z),\qquad\sigma(Z)^2= \beta ( Z - Z^2 ).
\end{equation}
This property of additivity is generic, and will be used extensively to construct the dual transition of the OTOC process.

\subsubsection{Particle jumping}
We now consider a particle system on two sites, 1 and 2, and the only process is the jump of particles from site 1 to site 2 with rate $G_{21} h_1$. The non-zero elements of the matrices $M_P$ and $M_Q$ are
\begin{equation}
\begin{aligned}
M_P\Big( \begin{spmatrix}n_1-1\\ n_2 + 1\end{spmatrix}, \begin{spmatrix}n_1\\ n_2\end{spmatrix}\Big) &= G_{21} n_1   & M_P\Big( \begin{spmatrix}n_1\\ n_2\end{spmatrix}, \begin{spmatrix}n_1\\ n_2\end{spmatrix}\Big) & = - G_{21} n_1 \\
M_Q\Big(\begin{spmatrix}n_1\\ n_2\end{spmatrix}, \begin{spmatrix}n_1-1\\ n_2 + 1\end{spmatrix} \Big) &= G_{21} n_1   & M_Q\Big( \begin{spmatrix}n_1\\ n_2\end{spmatrix}, \begin{spmatrix}n_1\\ n_2\end{spmatrix}\Big) & = - G_{21} n_1 
\end{aligned}
\end{equation}
so that
\begin{equation}
\partial_t Q(\vec n,t) =\sum_{\vec m} M_Q(\vec n,\vec m) Q(\vec m,t)=G_{21}n_1 \Big[Q\Big(\begin{spmatrix}n_1-1\\ n_2 + 1\end{spmatrix},t\Big)- Q(\vec n,t)\Big]=
G_{21}n_1 \bigg\langle  Z_1(t)^{n_1}Z_2(t)
^{n_2}\Big[\frac{Z_2(t)}{Z_1(t)}-1\Big]\bigg\rangle.
\end{equation}
Comparing with \autoref{eq:dQ}, we see that we must take
\begin{equation}
\mu_1=G_{21}(Z_2-Z_1),\quad\mu_2=0,\qquad\sigma_1=\sigma_2=0.
\end{equation}

\subsubsection{Birth at distance}\label{sec:goodprocess}
In this variant of the previous case, particles on site 1 provoke a birth at site 2, which means that the number of particles at site 2 increases by 1 with rate $G_{21} h_1$, without changing the value $h_1$ at site 1.
The non-zero elements of the matrices $M_P$ and $M_Q$ are
\begin{equation}
\begin{aligned}
M_P\Big( \begin{spmatrix}n_1\\ n_2 + 1\end{spmatrix}, \begin{spmatrix}n_1\\ n_2\end{spmatrix}\Big) &= G_{21} n_1   & M_P\Big( \begin{spmatrix}n_1\\ n_2\end{spmatrix}, \begin{spmatrix}n_1\\ n_2\end{spmatrix}\Big) & = - G_{21} n_1 \\
M_Q\Big(\begin{spmatrix}n_1\\ n_2\end{spmatrix}, \begin{spmatrix}n_1\\ n_2 + 1\end{spmatrix} \Big) &= G_{21} n_1   & M_Q\Big( \begin{spmatrix}n_1\\ n_2\end{spmatrix}, \begin{spmatrix}n_1\\ n_2\end{spmatrix}\Big) & = - G_{21} n_1 
\end{aligned}
\label{eq:MPQatdistance}
\end{equation}
so that
\begin{equation}
\partial_t Q(\vec n,t) =\sum_{\vec m} M_Q(\vec n,\vec m) Q(\vec m,t)=G_{21}n_1 \Big[Q\Big(\begin{spmatrix}n_1\\ n_2 + 1\end{spmatrix},t\Big)- Q(\vec n,t)\Big]=
G_{21}n_1 \bigg\langle  Z_1(t)^{n_1}Z_2(t)
^{n_2}\Big[Z_2(t)-1\Big]\bigg\rangle.
\label{eq:C33}
\end{equation}
Comparing with \autoref{eq:dQ}, we see that we must take
\begin{equation}
\mu_1=G_{21}(Z_1Z_2-Z_1)=G_{21}Z_1(Z_2-1),\quad\mu_2=0,\qquad\sigma_1=\sigma_2=0.
\end{equation}
The additional factor $Z_1$ in the first term distinguishes it from the jumping process. 
It is interesting to notice that, even though $h_1$ remains constant and $h_2$ increases in this situation, it is $Z_2$ which remains constant and $Z_1$ which varies in the dual process.

\subsubsection{Birth at distance with arrival-dependent rate and hard constraint}
\label{sec:badprocess}
In this variant, which is directly related to the OTOC model, there is a ``birth at distance'' from site~1 to~2, but with a rate $G_{21}h_1(1-\frac{h_2}{N})$ which is limited by the occupancy $h_2$ of site~2. In particular, the rate reaches 0 when $h_2=N$ ($N$ is some integer parameter). The non-zero elements of the matrices $M_P$ and $M_Q$ gain a factor $(1-\frac{n_2}{N})$ compared to \autoref{eq:MPQatdistance}, and \autoref{eq:C33}
becomes
\begin{equation}
\partial_t Q(\vec n,t) =
G_{21}n_1 \left(1-\frac{n_2}{N}\right) \bigg\langle  Z_1(t)^{n_1}Z_2(t)
^{n_2}\Big[Z_2(t)-1\Big]\bigg\rangle.
\end{equation}
We compare to \autoref{eq:dQ}, and see that we have a problem: there is no way to choose $\mu_1$, $\mu_2$, $\sigma_1$ and $\sigma_2$ in such a way that a cross term $n_1n_2$ appears.
There is no way to build a dual process in this case with our framework.

We could try to extend our framework, and allow some correlation in our Brownian processes: $\langle\diffd B_1 \diffd B_2\rangle = C_{12}\diffd t$ where $\diffd B_1$
 and $\diffd B_2$ are the increments of $B_1$ and $B_2$ during the same time interval $\diffd t$.
Then, \autoref{eq:dQ} becomes
\begin{equation}
\partial_tQ(\vec n,t)  = \bigg\langle Z_1(t)^{n_1}Z_2(t)^{n_2}\bigg(\sum_{i=1,2}   \Big[\frac{\mu_i n_i}{Z_i(t)}+\frac{\sigma_i^2 n_i(n_i-1)}{2Z_i(t)^2}\Big]+C_{12} \frac{\sigma_1\sigma_2n_1n_2}{Z_1(t)Z_2(t)}\bigg)\bigg\rangle.
\end{equation}
We do have a $n_1n_2$ term, but then we would need to have $\sigma_1$ and $\sigma_2$ non-zero, and that would bring in some unwanted terms $n_1^2$ and $n_2^2$. Adding correlations to the Brownian processes do not help us, and we are unable to find a dual process.

\subsection{Height Evolution of OTOC}
\subsubsection{On the lattice}

We now consider a $d$-dimensional lattice indexed by vector $\vec x$, and write $h(\vec x,t)$ for the height at position $\vec x$ and time $t$. The dual process will be made of one diffusion $Z(\vec x,t)$ per lattice site $\vec x$.

In the original ``hard constraint'' model, particles  make some births at distance with arrival-dependent rates, as in section~\ref{sec:badprocess}. As explained above, we are unable to construct a dual process to this kind of transitions.

For this reason we focus on the ``soft constraint'' model, where particles make some births at a distance with arrival-independent rates, as in section~\ref{sec:goodprocess}, with some on-site coalescence, as in section~\ref{sec:coalesce}. More specifically, the rate for a birth at site $\vec y$ from site $\vec x$ is assumed to be $G(\|\vec y-\vec x\|) h(\vec x,t)$, and the coalescence rate on site $\vec x$ is $\frac\beta2 h(\vec x,t)(h(\vec x,t)-1)$.

As already explained, we do not need to write all the transition matrices; we simply have to combine the results of sections~\ref{sec:goodprocess} and~\ref{sec:coalesce} to obtain the coupled SDE governing the dual processes $Z(\vec x,t)$; for all the sites $\vec x$, one has
\begin{equation}
\diffd Z(\vec x,t) = Z(\vec x,t)\sum_{\vec y} G(\|\vec y-\vec x \| )\big(Z(\vec y,t)-1\big) \,\diffd t
+\sqrt{\beta \big[Z(\vec x,t)-Z(\vec x,t)^2\big]} \diffd B(\vec x,t).
\label{C37}
\end{equation}
(Remark: the terms $G_{\vec y,\vec x}$ for $\vec y\ne\vec x$ represent ``birth at a distance'' process, while the term $G_{\vec x,\vec x}$ when $\vec y=\vec x$ represents   the ``birth on site'' process, as can be checked from section~\ref{sec:birthonsite}.)

\subsubsection{In the continuum limit}

We now restrict ourselves to one dimension of space.
\begin{align}
\diffd Z_i(t) &= Z_i(t)\sum_j G(|i-j|)\big(Z_j(t)-1\big) \,\diffd t
+\sqrt{\beta \big[Z_i(t)-Z_i(t)^2\big]} \diffd B_i(t). 
\\&= Z_i(t)\sum_j G(|i-j|)\big(Z_j(t)-Z_i(t)\big) \,\diffd t+\lambda\big(Z_i(t)^2-Z_i(t)\big)
+\sqrt{\beta \big[Z_i(t)-Z_i(t)^2\big]} \diffd B_i(t),
\label{eq:Z1d}
\end{align}
where $\lambda= \sum_j G(|i-j|)$.

We have $G(|i-j|)$ decays as $|i-j|^{-2\alpha}$ when $|i-j|$ is large. To have a valid thermodynamic limit, we take $\alpha > \frac{1}{2}$. A continuous version of the lattice equation \eqref{eq:Z1d} would look like 
\begin{equation}
\begin{cases}
\partial_t Z = Z \tilde D \Delta  Z+\tilde \lambda (Z^2-Z)+\sqrt{\tilde\beta (Z-Z^2)}\,\eta&\text{if $\alpha>\frac32$},\\
\partial_t Z = -Z \tilde D (-\Delta)^{\alpha-\frac{1}{2}} Z+\tilde \lambda (Z^2-Z)+\sqrt{\tilde\beta (Z-Z^2)}\,\eta&\text{if $\alpha<\frac32$},
\end{cases}
\end{equation}
where $\tilde D$, $\tilde\lambda$ and $\tilde\beta$ are rescaled coefficients and where $\eta(x,t)$ is a space and time white noise: $\langle \eta(x,t)\eta(x',t')\rangle=\delta(x-x')\delta(y-y')$.

In terms of $f = 1- Z$, we would have (for $\frac12<\alpha<\frac32$)
\begin{equation}
  \partial_t f = -(1-f) \tilde D ( - \Delta)^{\alpha - \frac{1}{2}} f + \tilde\lambda f (1 - f) + \sqrt{\tilde \beta f (1 - f) }\, \eta
\label{eq:C.f}
\end{equation}
which is \autoref{eq:noisy_fkpp} in the main text. 

\section{Scaling function of $\ell(t)$ close to $\alpha = 1$ in 1D}

In this appendix, we derive  \autoref{eq:criticalell} and give a more general scaling form for $\ell(t)$ when $\delta = 2(\alpha - 1)$ is small.

\subsection{Main equation}
We start from \autoref{eq:iter_equ_2} in dimension $d=1$, which is assumed to be valid for large times. We write $2\alpha=2+\delta$ and, for later convenience, we introduce $c$ by $K=2^{-c}/\sqrt\pi$. Thus, our starting point is
\begin{equation}
\ell(t)^{2+\delta} = \frac{2^{-c}}{\sqrt\pi} \int_0^t \ell(\tau) \ell(t-\tau)\,\diffd \tau.
\label{eq:startingpoint}
\end{equation}
The goal is to study the behaviour of $\ell(t)$ close to the transition, when $|\delta|\ll1$. 
The value of $c$ can be estimated from Fig.~\ref{fig:ell(t)}(c), which gives $\frac{1}{K} = \sqrt\pi\, 2^c   \simeq0.04$  for $\alpha=1$ (i.e.\@ $\delta=0$). Note that, strictly speaking, $c$ depends on $\delta$, but it is clear from Fig.~\ref{fig:ell(t)}(c) that we have $c\simeq c(0)$ for $\delta$ small enough.

As $\ell(t)$ increases quickly with time, we expect the integral in \autoref{eq:startingpoint}
 to be dominated by a small interval of $\tau$ around $t/2$, and we make the change of varitable $\tau=t/2+ut$:
\begin{equation}
\ell(t)^{2+\delta} = \frac{2^{-c}}{\sqrt\pi}\, t \int_{-\frac12}^{\frac12} \ell\bigl(\tfrac t2 + u t\bigr)\ell\bigl(\tfrac t2 - u t\bigr) \,\diffd u.
\end{equation}
We use logarithmic variables:
\begin{equation}
\varphi(z,\delta)=\log_2\ell(t)\quad\text{with } z =\log_2 t\qquad\iff\qquad
\ell(t)=2^{\varphi(\log_2 t,\delta)}.
\end{equation}
Then, $\log_2(\frac t2\pm ut)=z-1+\log_2(1\pm 2u)$ and we obtain
\begin{equation}
2^{(2+\delta)\varphi(z,\delta)}=\frac{2^{-c}}{\sqrt\pi}\, 2^z \int_{-\frac12}^{\frac12} 2^{\varphi\big(z-1+\log_2(1+2u),\delta\big)+\varphi\big(z-1+\log_2(1-2u),\delta\big)}
\,\diffd u.
\label{eq:main_with_phi}
\end{equation}

\subsection{The scaling limit}
We now assume that $\delta$ is small but non-zero, and that, simultaneously, $z=\log_2t$ is large.
As shown in \cite {hallatschek_acceleration_2014}, the correct scaling is
\begin{equation}
\varphi(z,\delta) = \frac1{\delta^2}f(y,\delta)\quad\text{with }y=\frac{\delta z}2
\label{eq:phi_vs_f}
\end{equation}
where the function $f(y,\delta)$ remains of order 1 when $\delta$ goes to zero while $y$ remains fixed. To leading order, \cite {hallatschek_acceleration_2014} found $f(y,\delta)=2[\exp(-y)+y-1]$. We are going to show this again and obtain next order corrections.
In what follows, we write $f'(y,\delta)$, $f''(y,\delta)$, etc.\@ to mean derivatives of $f$ with respect to $y$.

The main equation \autoref{eq:main_with_phi} in terms of $f$ becomes
\begin{equation}
2^{\frac{2+\delta}{\delta^2}f(y,\delta)}=\frac{2^{-c}}{\sqrt\pi}\, 2^{\frac {2y}{\delta}}
 \int_{-\frac12}^{\frac12} 2^{X(y,\delta,u)}\,\diffd u
\label{eq:main_with_f}
 \end{equation}
 with
 \begin{equation}
X(y,\delta,u)=
\frac1{\delta^2}f\left(y-\tfrac\delta2+\tfrac\delta2\log_2(1+2u),\delta\right)+  \frac1{\delta^2}f\left(y-\tfrac\delta2+\tfrac\delta2\log_2(1-2u),\delta\right).
\end{equation}

Let us expand $X(y,\delta,u)$ for small $\delta$; we obtain, to leading order,
\begin{equation}
X(y,\delta,u)\simeq
\frac2{\delta^2}f(y,\delta)+\frac1{2\delta}f'(y,\delta)\bigl(-2+\log_2(1-4u^2)\bigr).
\label{eq:expansion1}
\end{equation}
First, we can see here that the scaling \autoref{eq:phi_vs_f} is correct. Indeed, the leading term $\frac2{\delta^2}f(y,\delta)$ in \autoref{eq:expansion1} cancel with the same term in the left hand side of \autoref{eq:main_with_f}; then come three terms of the same order:  $\frac1\delta f(y,\delta)$ on the left hand side, $\frac{2y}\delta$ in front of the integral and a term proportional to $\frac1\delta f'(y,\delta)$ in the integral, as seen in \autoref{eq:expansion1}. Had we taken a more general scaling function  $\varphi(z,\delta)=\frac1{\delta ^p} f(\frac{1}{2}\delta^q z,\delta)$ for some $p$ and $q$, we would have found those three terms to be, respectively, of order $\delta^{1-p}$, $\delta^{-q}$ and $\delta^{q-p}$. The only way to have these three terms of the same order of magnitude is indeed to take $p=2$ and 
$q=1$, as in  \autoref{eq:phi_vs_f}.

For fixed $y$ and small $\delta$, the function $f(y,\delta)$ is of order~1, and so is its derivative $f'(y,\delta)$. Furthermore, $\ell(t)$ increases with $t$, so $\varphi(z,\delta)$ increases with $z$ and $f(y,\delta)$ increases with $y$. We thus see that the factor $\frac1{2\delta} f'(y,\delta)$ in \autoref{eq:expansion1} is a large positive number when $\delta$ is small and, therefore, the integral in \autoref{eq:main_with_f} is dominated by values of $u$ which maximize $\log_2(1-4u^2)$, \textit{i.e.\@} for $u$ small. 
Then, writing $\log_2(1-4u^2)\simeq-\frac4{\ln2}u^2$ shows that the integral in \autoref{eq:main_with_f} is in fact dominated by values of $u$  of order     $\sqrt\delta$, so that  $u^2/\delta$ is of order 1.

With this in mind, we can now write a better expansion of $X(y,\delta,u)$, but we need to expand  at the same time in powers of $\delta$ and of $u$ (of order $\sqrt\delta$). In the hand, we will need to know $X(y,\delta,u)$ up to order $\delta$. Then the arguments of $f$ have to be known up to order $\delta^3$, and we must expand the logarithms up to $u^4$ and write $y-\frac\delta2+\frac\delta2\log_2(1\pm2u)=y-\frac\delta2+\frac1{\ln2}\big[\pm\delta u-\delta u^2\pm\tfrac43\delta u^3-2\delta u^4\big]+\cdots$.
We then obtain, after elementary algebra,
\begin{equation}
X(y,\delta,u)=\frac2{\delta^2}f(y,\delta)-\frac1{\delta}f'(y,\delta)\Bigl(1+\frac2{\ln2}u^2+\frac{4}{\ln2}u^4\Bigr)
+\frac14f''(y,\delta)\Bigl(1+\frac{4+4\ln2}{\ln^22}u^2\Bigr)-\frac{\delta}{24}f'''(y,\delta)+o(\delta).
\label{eq:expansion3}
\end{equation}
Due to the long form of the equations, in the following we will omit $o(\delta)$, equal signs should be understood as ``equal up to $\delta$" when necessary from the context. 

We plug \autoref{eq:expansion3} into \autoref{eq:main_with_f} to obtain, after small simplifications,
\begin{align}
2^{\frac{1}{\delta}f(y,\delta)}
&=\frac{2^{-c}}{\sqrt\pi}\, 2^{\frac {2y}{\delta}-\frac1\delta f'(y,\delta)+\frac14f''(y,\delta)-\frac{\delta}{24}f'''(y,\delta)}
 \int_{-\frac12}^{\frac12} 2^{
-\frac2{\delta\ln2}f'(y,\delta)(u^2+2u^4)
+\frac{1+\ln2}{\ln^22}f''(y,\delta)u^2}\,\diffd u,
\\&=\frac{2^{-c}}{\sqrt\pi}\, 2^{\frac {2y}{\delta}-\frac1\delta f'(y,\delta)+\frac14f''(y,\delta)-\frac{\delta}{24}f'''(y,\delta)}
 \int_{-\frac12}^{\frac12} e^{
-\frac2{\delta}f'(y,\delta)u^2-\frac{4}{\delta}f'(y,\delta)u^4
+\frac{1+\ln2}{\ln2}f''(y,\delta)u^2}\,\diffd u
.
\end{align}
where we simply wrote $2^X=e^{X\ln2}$ in the last step. In the exponential function of the integral, the term $-\frac2{\delta}f'(y,\delta)u^2$ is of order one, while the two other terms are of order $\delta$. We expand them to obtain
\begin{equation}
2^{\frac{1}{\delta}f(y,\delta)}
=\frac{2^{-c}}{\sqrt\pi}\, 2^{\frac {2y}{\delta}-\frac1\delta f'(y,\delta)+\frac14f''(y,\delta)-             \frac{\delta}{24}f'''(y,\delta)}
 \int_{-\frac12}^{\frac12} e^{
-\frac2{\delta}f'(y,\delta)u^2}\biggl[1-\frac{4}{\delta}f'(y,\delta)u^4
+\frac{1+\ln2}{\ln2}f''(y,\delta)u^2\biggr]\diffd u.
\end{equation}
The domain of integration can be pushed to $(-\infty,\infty)$ up to exponentially small terms. Then applying Gaussian integrals $\int e^{-a u^2}\diffd u=\sqrt{\pi/a}$, $\int e^{- a u^2}u^2\,\diffd u=\sqrt{\pi/a}\times\frac1{2a}$ and
 $\int e^{- a u^2}u^4\,\diffd u=\sqrt{\pi/a}\times\frac3{4a^2}$ gives
\begin{equation}
2^{\frac{1}{\delta}f(y,\delta)}
=\frac{2^{-c}}{\sqrt\pi}\, 2^{\frac {2y}{\delta}-\frac1\delta f'(y,\delta)+\frac14f''(y,\delta)- \frac{\delta}{24}f'''(y,\delta)}
\sqrt{\frac{\pi\delta}{2f'(y,\delta)}}
\biggl[1-\frac{3}{4f'(y,\delta)}\delta
+\frac{1+\ln2}{4\ln2}\frac{f''(y,\delta)}{f'(y,\delta)}\,\delta\biggr].
\end{equation}
The $\sqrt\pi$ cancels; we take the $\log_2$ of both sides and expand up to order $\delta$.
\begin{equation}
\frac{1}{\delta}\bigl[f(y,\delta)+f'(y,\delta)\bigr]=\frac {2y}{\delta}+\frac14f''(y,\delta)-c-                \frac{\delta}{24}f'''(y,\delta)-\frac12\log_2\frac{2f'(y,\delta)}\delta -\frac{3}{4f'(y,\delta)\ln2}\delta
+\frac{1+\ln2}{4\ln^22}\frac{f''(y,\delta)}{f'(y,\delta)}\delta.
\label{eq:main_f}
\end{equation}
(This equality and the ones before are understood to be ``up to order $\delta$''.) 
We now expand $f$ in powers of $\delta$:
\begin{equation}
f(y,\delta)=f_0(y)+\delta f_1(y,\delta)+\delta^2f_2(y,\delta)+\cdots
\label{eq:f_expansion}
\end{equation}
Note that we still kept a dependency on $\delta$ for $f_1$ and $f_2$; indeed, it will turn out that they must depend logarithmically on $\delta$.

We plug \autoref{eq:f_expansion} into \autoref{eq:main_f} and collect the terms of different orders in $\delta$; the only non-trivial term is the $\log_2$ term which gives:
\begin{equation}
\log_2\frac{2f'(y,\delta)}\delta
=\log_2\Bigl(\frac{2f'_0(y)\big(1+\delta \frac{f_1'(y,\delta)}{f'_0(y)}\big)+\cdots}\delta\Bigr)
=\log_2\frac{2f_0'(y)}\delta+\frac{\delta}{\ln2} \frac{f_1'(y,\delta)}{f_0'(y)}+\cdots
\end{equation}
We finally arrive at
\begin{equation}
\begin{cases}
\displaystyle f_0(y)+f_0'(y) = 2y,\\[1ex]
\displaystyle f_1(y,\delta)+f_1'(y,\delta) = \frac14f_0''(y)-c-\frac12\log_2\frac{2f_0'(y)}\delta ,\\[1ex]
\displaystyle f_2(y,\delta)+f_2'(y,\delta) = \frac14f_1''(y,\delta)-\frac1{24}f_0'''(y)-\frac1{2\ln2}\frac{f_1'(y,\delta)}{f_0'(y)}-            \frac{3}{4f'_0(y)\ln2} +\frac{1+\ln2}{4\ln^22}\frac{f_0''(y)}{f_0'(y)}.
\end{cases}
\label{eq:edpf_i}
\end{equation}
We will solve these equations order by order. While doing so, we must remember that $\varphi(z,\delta)$ has a limit when $\delta\to0$ with $z$ fixed, while $\varphi(z,\delta)=\frac1{\delta^2}f(\frac{\delta z}2,\delta)=\frac1{\delta^2}f_0(\frac{\delta z}2)+\frac1{\delta}f_1(\frac{\delta z}2,\delta)+f_2(\frac{\delta z}2,\delta)+\cdots$, and so
\begin{equation}
\frac1{\delta^2}f_0\Bigl(\frac{\delta z}2\Bigr), \quad
\frac1{\delta}  f_1\Bigl(\frac{\delta z}2,\delta\Bigr)\quad\text{and}\quad
  f_2\Bigl(\frac{\delta z}2,\delta\Bigr)\quad\text{have limits when $\delta\to0$.}
\label{eq:constraintf_i}
\end{equation}

We now solve line by line \autoref{eq:edpf_i} with the constraint \autoref{eq:constraintf_i}. The general solution for $f_0$ would be $f_0(y)=Ae^{-y}+2y-2$, but with \autoref{eq:constraintf_i} one must take $A=2$. Finally
\begin{equation}
f_0(y)=2(e^{-y}+y-1).
\label{eq:f0}
\end{equation}
This is essentially the leading order scaling derived in Ref.~\cite{hallatschek_acceleration_2014}.
We plug this in the second line of  \autoref{eq:edpf_i} to obtain
\begin{equation}
f_1(y,\delta)+f_1'(y,\delta) = \frac12e^{-y}-c-1-\frac12\log_2\frac{1-e^{-y}}\delta.
\end{equation}
The solution satisfying \autoref{eq:constraintf_i} is
\begin{equation}
f_1(y,\delta) =\Bigl(\frac12+\frac1{2\ln2}\Bigr)ye^{-y} -\frac12(1-e^{-y})\log_2\frac{1-e^{-y}}\delta
-(c+1)(1-e^{-y}).
\label{eq:f1}
\end{equation}
We will need $f_1'(y,\delta)$:
\begin{equation}
\begin{aligned}
f_1'(y,\delta)&=-\Bigl(\frac12+\frac1{2\ln2}\Bigr)ye^{-y}-\big(c+\tfrac12\big)e^{-y}-\frac12e^{-y}\log_2\frac{1-e^{-y}}\delta,
\\&=-\frac{1}{2}ye^{-y}-\big(c+\tfrac12\big)e^{-y}-\frac12e^{-y}\log_2\frac{e^{y}-1}\delta.
\end{aligned}
\label{eq:f1'}
\end{equation}

We now turn to $f_2$. Multiplying the third line of  \autoref{eq:edpf_i} by $e^y$ and integrating leads to
\begin{equation}
f_2(y,\delta)e^y =C+\int\diffd y\, e^y\biggl[ \frac14f_1''(y,\delta)-\frac1{24}f_0'''(y)-\frac1{2\ln2}\frac{f_1'(y,\delta)}{f_0'(y)}-            \frac{3}{4f'_0(y)\ln2}
+\frac{1+\ln2}{4\ln^22}\frac{f_0''(y)}{f_0'(y)}
\biggr],
\label{eq:f2asint}
\end{equation}
where $C$ is an arbitrary constant.
Let us go term by term; in what follows, everything is ``up to an additive constant'' which is folded into $C$.
\begin{description}
\item[First term in \autoref{eq:f2asint}] We have, using \autoref{eq:f1'}
\begin{equation}
\begin{aligned}
\int\diffd y\, e^{y}f''_1(y,\delta)&=e^{y}f_1'(y,\delta)-\int\diffd y\, e^{y}f'_1(y,\delta))
,
\\&
=-\frac{1}{2}y-\frac12\log_2\frac{e^{y}-1}\delta +\int\diffd y \biggl[\frac{1}{2}y+\big(c+\tfrac12\big)+\frac12\log_2\frac{e^{y}-1}\delta\biggr]
\\&
=\frac14y^2+cy-\frac12\log_2\frac{e^{y}-1}\delta+\frac12\int\diffd y\,\log_2\frac{e^{y}-1}\delta
.
\end{aligned}
\end{equation}
Notice that we droped the constant $c+\frac12$.
There remains a term with a complicated primitive. Let 
\begin{equation}
Q(y,\delta)=\int_0^y\diffd y'\,\log_2\frac{e^{y'}-1}\delta.
\label{eq:defQ}
\end{equation}

\item[Second term in \autoref{eq:f2asint}] We have, from \autoref{eq:f0}, $f_0'''(y)=-2e^{-y}$. Hence
\begin{equation}
\int\diffd y\, e^{y}f'''_0(y)=-2y.
\end{equation}

\item[Third term in \autoref{eq:f2asint}] One has, $f_0'(y)=2(1-e^{-y})$ and so 
$\frac1{f_0'(y)}=\frac12\frac{e^y}{e^y-1}$. With \autoref{eq:f1'},
\begin{equation}
e^{y}\frac{f_1'(y,\delta)}{f_0'(y)}=
-\frac{1}{4}\frac {ye^y}{e^{y}-1}-\Bigl(\frac c2+\frac14\Bigr)\frac{e^y}{e^{y}-1}-\frac14\frac{e^y\log_2\frac{e^{y}-1}\delta}{e^{y}-1}.
\label{eq:thirdterm1}
\end{equation}
We now integrate \autoref{eq:thirdterm1}.
The first term needs to be integrated by parts:
\begin{equation}
\int\diffd y\,
\frac {ye^y}{e^{y}-1}=
y \ln\frac{e^{y}-1}\delta -
\int\diffd y \,\ln\frac{e^{y}-1}\delta
=
(\ln2)y \log_2\frac{e^{y}-1}\delta -(\ln 2) Q(y,\delta),
\end{equation}
where $Q(y,\delta)$ was defined in \autoref{eq:defQ}.
Notice that, when integrating $e^y/(e^y-1)$, we chose for primitive $\ln\frac{e^y-1}\delta$ in order to obtain a function which has a $\delta\to0$ limit when $y=z\delta/2$. We shall do the same regularization several times below. 

The two other terms in \autoref{eq:thirdterm1} can be directly integrated, and we obtain
\begin{equation}
\int\diffd y\,e^{y}\frac{f_1'(y,\delta)}{f_0'(y)}=
-\frac{\ln2}{4}y \log_2\frac{e^{y}-1}\delta +
 \frac{\ln2}{4}Q(y,\delta)
 -\Bigl(\frac c2+\frac14\Bigr)(\ln2)\log_2\frac{e^{y}-1}\delta-\frac{\ln2}8\biggl(\log_2\frac{e^{y}-1}\delta\biggr)^2.
\label{eq:thirdterm}
\end{equation}

\item[Fourth term in \autoref{eq:f2asint}] One has
\begin{equation}
\int\diffd y  \frac{e^{y}}{f_0'(y)}
= \frac12\int\diffd y \frac{e^{2y}}{e^{y}-1}
=  \frac12\int\diffd y\biggl[e^y+\frac{e^{y}}{e^{y}-1}\biggr]
= \frac12\biggl[e^y+
(\ln2) \log_2\frac{e^{y}-1}\delta\biggr].
\end{equation}

\item[Fifth and last term in \autoref{eq:f2asint}] One has
\begin{equation}
\int\diffd y\, e^{y} \frac{f_0''(y)}{f_0'(y)} =
\int\diffd y\,\frac{e^y}{e^{y}-1} 
=(\ln2) \log_2\frac{e^{y}-1}\delta
.
\end{equation}
\end{description}

Putting everything together, we arrive at
\begin{equation}
\begin{aligned}
f_2(y,\delta)e^y &=C+\frac14\biggl[\frac14y^2+cy-\frac12\log_2\frac{e^{y}-1}\delta+\frac12Q(y,\delta)\biggr]
-\frac1{24}\biggl[-2y\biggr]\\
&\quad -\frac1{2\ln2}\biggl[
-\frac{\ln2}{4}y \log_2\frac{e^{y}-1}\delta +
 \frac{\ln2}{4}Q(y,\delta)
 -\Bigl(\frac c2+\frac14\Bigr)(\ln2)\log_2\frac{e^{y}-1}\delta-\frac{\ln2}8\Bigl(\log_2\frac{e^{y}-1}\delta\Bigr)^2
\biggr]
\\&\quad
-  \frac{3}{4\ln2}\times
\frac12\biggl[e^y+
(\ln2) \log_2\frac{e^{y}-1}\delta\biggr]
+\frac{1+\ln2}{4\ln^22}\biggl[(\ln2) \log_2\frac{e^{y}-1}\delta
\biggr],
\\&=C+\frac{1}{16}y^2
+\Bigl(\frac1{12}+\frac c4\Bigr)y
-\frac3{8\ln2}e^y
\\&\qquad
+\frac{1}{8}y\log_2\frac{e^{y}-1}\delta
+\Bigl(\frac1{4\ln2}-\frac18+\frac{c}4\Bigr)\log_2\frac{e^{y}-1}\delta
+\frac1{16}
\Bigl(\log_2\frac{e^{y}-1}\delta\Bigr)^2,
\\&=C+\frac1{16}\Bigl(y+\log_2\frac{e^{y}-1}\delta\Bigr)^2
+\Bigl(\frac5{24}-\frac1{4\ln2}\Bigr)y
-\frac3{8\ln2}e^y
+\left(\frac1{4\ln2}-\frac18+\frac{c}4\right)\Bigl(y+\log_2\frac{e^{y}-1}\delta
\Bigr)
\\&=C+\frac1{16}\Bigl(\frac2{\ln2}-1+2c+y+\log_2\frac{e^{y}-1}\delta\Bigr)^2
+\Bigl(\frac5{24}-\frac1{4\ln2}\Bigr)y
-\frac3{8\ln2}e^y,
\end{aligned}
\label{eq:f2_1}
\end{equation}
where we changed the definition of $C$ changed in the last line.
Notice that the two $Q(y,\delta)$ have cancelled. Finally
\begin{equation}
f_2(y,\delta)=
Ce^{-y}+\frac1{16}\Bigl(\frac2{\ln2}-1+2c+y+\log_2\frac{e^{y}-1}\delta\Bigr)^2e^{-y}
+\left(\frac5{24}-\frac1{4\ln2}\right)ye^{-y}
-\frac3{8\ln2}
.
\label{eq:f2}
\end{equation}

From \autoref{eq:phi_vs_f}, \autoref{eq:f_expansion}, \autoref{eq:f0}, \autoref{eq:f1} and \autoref{eq:f2},
our scaling function is therefore, up to vanishing terms,
\begin{equation}
\begin{aligned}
\varphi(z,\delta)&=\frac2{\delta^2}\bigl(e^{-y}+y-1\bigr)
+\frac1\delta\biggl[\Bigl(\frac12+\frac1{2\ln2}\Bigr)ye^{-y} -\frac12(1-e^{-y})\log_2\frac{1-e^{-y}}\delta
-(c+1)(1-e^{-y})\biggr]
\\&\quad
+ Ce^{-y}
+\frac1{16}\Bigl(\frac2{\ln2}-1+2c+y+\log_2\frac{e^{y}-1}\delta\Bigr)^2e^{-y}
+\Bigl(\frac5{24}-\frac1{4\ln2}\Bigr)ye^{-y}
-\frac3{8\ln2}+\cdots
\end{aligned}
\label{eq:scaling}
\end{equation}
where we recall that $c$ is defined in \autoref{eq:startingpoint},
that $C$ is an unknown constant depending on the details of the model, and that
\begin{equation}
\delta=2\alpha-2,\qquad
z=\log_2t,\qquad
\varphi(z,\delta)=\log_2\ell(t),\qquad
y= \frac{z\delta}2.
\end{equation}

\subsection{The critical case}
The critical case $\alpha=1$, or $\delta=0$ can be recovered from \autoref{eq:scaling} by taking the limit $\delta\to0$ after replacing $y$ by $z\delta/2$. One obtains
\begin{equation}
\begin{aligned}
\varphi(z,0)&=\frac{z^2}4+
\Bigl[\Bigl(\frac14+\frac1{4\ln2}\Bigr) -\frac 14\log_2\frac z2
-\frac 1 2(c+1)\Bigr]z
+\Bigl[
C+\frac1{16}\Bigl(\frac2{\ln2}-1+2c+\log_2\frac z2\Bigr)^2
-\frac3{8\ln2}\Bigr]+\cdots
\\&=
\frac{z^2}4+
\Bigl[ -\frac 14\log_2 z+\frac1{4\ln2}-\frac c2
\Bigr]z
+\Bigl[
\frac1{16}\Bigl(\log_2 z+\frac2{\ln2}-2+2c\Bigr)^2
+C-\frac3{8\ln2}\Bigr]+\cdots
\end{aligned}
\end{equation}
The different terms are of order $z^2$, $z$, $1$, etc.\@ (with logarithmic corrections).
We can rearrange terms to obtain, with a different constant $C$,
\begin{equation}
\varphi(z,0)=\frac{1}4\Bigl(z -\frac 12\log_2 z+\frac1{2\ln2}-c\Bigr)^2
+ \Bigl(\frac3{8\ln2}-\frac14\Bigr)\log_2z + C+\cdots
\label{eq:scaling0}
\end{equation}
Going back to $\ell(t)$ and $t$, we obtain
\begin{equation}
\log_2\ell(t)=\frac{1}4\Bigl(\log_2t -\frac 12\log_2 \log_2t+\frac1{2\ln2}-c\Bigr)^2
+ \Bigl(\frac3{8\ln2}-\frac14\Bigr)\log_2\log_2 t + C+\cdots
\end{equation}
or, with yet another constant $C$,
\begin{equation}
\ln\ell(t)=\frac{1}{4\ln2}\Bigl(\ln t -\frac 12\ln \ln t+\frac12\ln\ln2+\frac1{2}-c\ln2\Bigr)^2
+ \Bigl(\frac3{8\ln2}-\frac14\Bigr)\ln\ln t + C+\cdots
\end{equation}
The constant term in the first parenthesis could be expressed as $\frac12+\frac12\ln\ln2+\ln(K\sqrt\pi)$. To be compared to \autoref{eq:criticalell}.

Note that these expression at $\delta=0$ can be obtained more directly from 
 \autoref{eq:main_with_phi}
without going through the resolution of the scaling function. 

\subsection{Numerical checks}
\begin{figure}[ht]
\includegraphics[width=.439\textwidth]{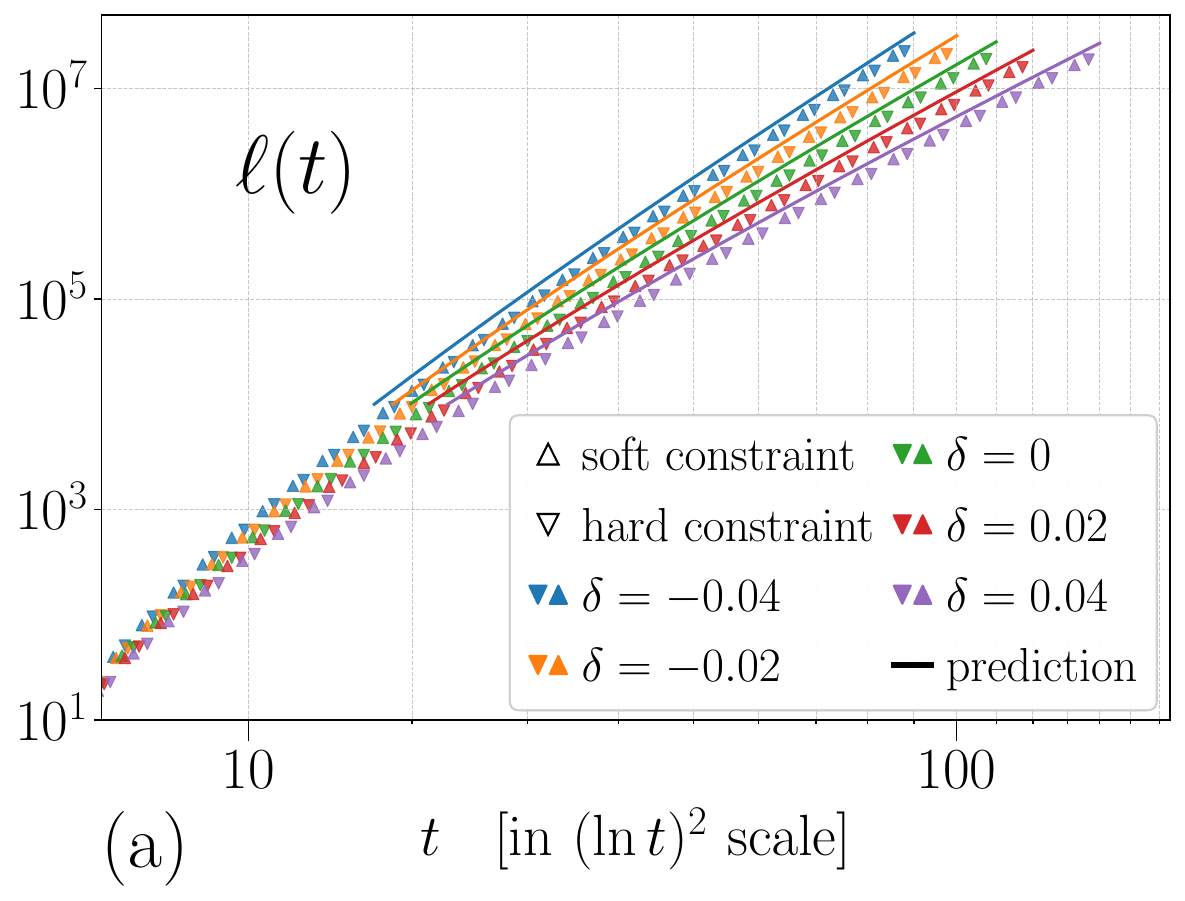}
\includegraphics[width=.461\textwidth]{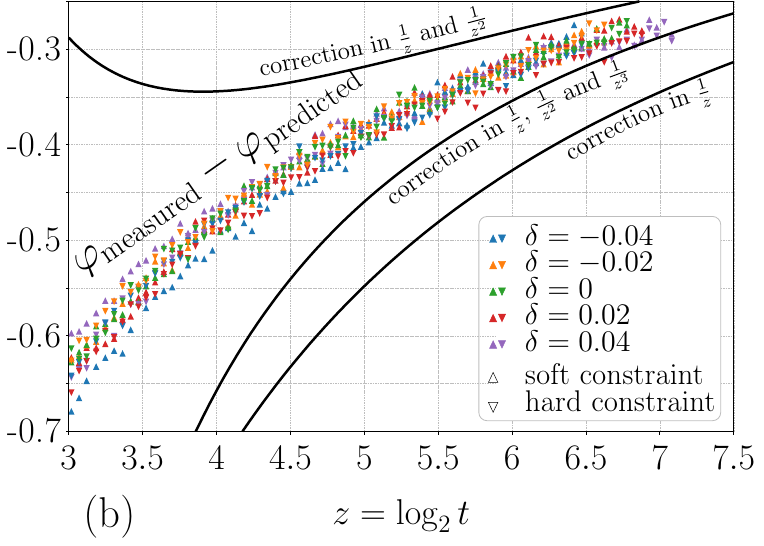}
\caption{Simulations of the soft constraint and hard constraint models for $\delta$ close to 0. (a) $\ell(t)$ as a function of $t$, and the prediction of the soft constraint model derived from \eqref{eq:scaling} (for $\delta\ne0$) or \eqref{eq:scaling0} (for $\delta=0$), as a function of $t$ (in $\ln^2t$ scale).  The prediction for the hard constraint model is very close (now shown). 
(b) $\varphi=\log_2\ell$ minus the prediction \eqref{eq:scaling} or \eqref{eq:scaling0}, as a function of $z=\log_2t$. The three black lines are the prediction for the vanishing corrections up to one, two or three orders, in the case $\delta=0$.}
\label{fig:scaling} 
\end{figure}

We compare the results \autoref{eq:scaling} and \autoref{eq:scaling0} to numerical simulation of the model with $\delta$ close to zero, in \autoref{fig:scaling}(a). Because of the quick spreading of the system, we are only able to simulate the system up to $t$ around 100 (depending on the value of $\delta$), which corresponds to $z$ being less than 7. If we were to continue the expansion \autoref{eq:scaling0}, we would get extra terms of order $1/z$, $1/z^2$, $1/z^3$, etc. (with logarithmic corrections). These extra terms vanish when $z\to\infty$, but they are still noticeable at $z=7$. We have computed these extra terms (but have not included them in the paper); in \autoref{fig:scaling}(b), we compare them to the difference between $\varphi=\log_2\ell(t)$ as measured, and the scaling results \autoref{eq:scaling} and \autoref{eq:scaling0}. Note that we only show the theoretical extra terms for $\delta=0$; we assume that they look similar for the other values of $\delta$ shown in this graph, which are all small.

To make these figures, we needed values of $c$ and of $C$ for each of the ten data-points (five values of $\delta$ in two models). The values of $c$ were obtained as in \autoref{fig:ell(t)}(c). The values of $C$ were actually hand-tuned so that the curves in \autoref{fig:scaling}(b) fall together between the black lines corresponding to two-order and three-order corrections.

\let\oldaddcontentsline\addcontentsline
\renewcommand{\addcontentsline}[3]{}
\bibliographystyle{apsrev4-1}

\bibliography{model_1_coal_paper}

\begin{thebibliography}{42}%
\makeatletter
\providecommand \@ifxundefined [1]{%
 \@ifx{#1\undefined}
}%
\providecommand \@ifnum [1]{%
 \ifnum #1\expandafter \@firstoftwo
 \else \expandafter \@secondoftwo
 \fi
}%
\providecommand \@ifx [1]{%
 \ifx #1\expandafter \@firstoftwo
 \else \expandafter \@secondoftwo
 \fi
}%
\providecommand \natexlab [1]{#1}%
\providecommand \enquote  [1]{``#1''}%
\providecommand \bibnamefont  [1]{#1}%
\providecommand \bibfnamefont [1]{#1}%
\providecommand \citenamefont [1]{#1}%
\providecommand \href@noop [0]{\@secondoftwo}%
\providecommand \href [0]{\begingroup \@sanitize@url \@href}%
\providecommand \@href[1]{\@@startlink{#1}\@@href}%
\providecommand \@@href[1]{\endgroup#1\@@endlink}%
\providecommand \@sanitize@url [0]{\catcode `\\12\catcode `\$12\catcode `\&12\catcode `\#12\catcode `\^12\catcode `\_12\catcode `\%12\relax}%
\providecommand \@@startlink[1]{}%
\providecommand \@@endlink[0]{}%
\providecommand \url  [0]{\begingroup\@sanitize@url \@url }%
\providecommand \@url [1]{\endgroup\@href {#1}{\urlprefix }}%
\providecommand \urlprefix  [0]{URL }%
\providecommand \Eprint [0]{\href }%
\providecommand \doibase [0]{http://dx.doi.org/}%
\providecommand \selectlanguage [0]{\@gobble}%
\providecommand \bibinfo  [0]{\@secondoftwo}%
\providecommand \bibfield  [0]{\@secondoftwo}%
\providecommand \translation [1]{[#1]}%
\providecommand \BibitemOpen [0]{}%
\providecommand \bibitemStop [0]{}%
\providecommand \bibitemNoStop [0]{.\EOS\space}%
\providecommand \EOS [0]{\spacefactor3000\relax}%
\providecommand \BibitemShut  [1]{\csname bibitem#1\endcsname}%
\let\auto@bib@innerbib\@empty
\bibitem [{\citenamefont {Larkin}\ and\ \citenamefont {Ovchinnikov}(1969)}]{larkin_quasiclassical_1969}%
  \BibitemOpen
  \bibfield  {author} {\bibinfo {author} {\bibfnamefont {A.~I.}\ \bibnamefont {Larkin}}\ and\ \bibinfo {author} {\bibfnamefont {Y.~N.}\ \bibnamefont {Ovchinnikov}},\ }\href@noop {} {\bibfield  {journal} {\bibinfo  {journal} {Soviet Journal of Experimental and Theoretical Physics}\ }\textbf {\bibinfo {volume} {28}},\ \bibinfo {pages} {1200} (\bibinfo {year} {1969})}\BibitemShut {NoStop}%
\bibitem [{\citenamefont {Shenker}\ and\ \citenamefont {Stanford}(2014)}]{shenker_black_2014}%
  \BibitemOpen
  \bibfield  {author} {\bibinfo {author} {\bibfnamefont {S.~H.}\ \bibnamefont {Shenker}}\ and\ \bibinfo {author} {\bibfnamefont {D.}~\bibnamefont {Stanford}},\ }\href {\doibase 10.1007/jhep03(2014)067} {\bibfield  {journal} {\bibinfo  {journal} {J. High Energ. Phys.}\ }\textbf {\bibinfo {volume} {2014}},\ \bibinfo {pages} {67} (\bibinfo {year} {2014})}\BibitemShut {NoStop}%
\bibitem [{\citenamefont {Roberts}\ and\ \citenamefont {Stanford}(2015)}]{roberts_diagnosing_2015}%
  \BibitemOpen
  \bibfield  {author} {\bibinfo {author} {\bibfnamefont {D.~A.}\ \bibnamefont {Roberts}}\ and\ \bibinfo {author} {\bibfnamefont {D.}~\bibnamefont {Stanford}},\ }\href {\doibase 10.1103/PhysRevLett.115.131603} {\bibfield  {journal} {\bibinfo  {journal} {Physical Review Letters}\ }\textbf {\bibinfo {volume} {115}},\ \bibinfo {pages} {131603} (\bibinfo {year} {2015})}\BibitemShut {NoStop}%
\bibitem [{\citenamefont {Kitaev}(2015)}]{kitaev2015}%
  \BibitemOpen
  \bibfield  {author} {\bibinfo {author} {\bibfnamefont {A.}~\bibnamefont {Kitaev}},\ }\href@noop {} {} (\bibinfo {year} {2015}),\ \bibinfo {note} {talks at KITP, April 7, 2015 and May 27, 2015}\BibitemShut {NoStop}%
\bibitem [{\citenamefont {Maldacena}\ \emph {et~al.}(2016)\citenamefont {Maldacena}, \citenamefont {Shenker},\ and\ \citenamefont {Stanford}}]{maldacena_bound_2015}%
  \BibitemOpen
  \bibfield  {author} {\bibinfo {author} {\bibfnamefont {J.}~\bibnamefont {Maldacena}}, \bibinfo {author} {\bibfnamefont {S.~H.}\ \bibnamefont {Shenker}}, \ and\ \bibinfo {author} {\bibfnamefont {D.}~\bibnamefont {Stanford}},\ }\href {\doibase 10.1007/jhep08(2016)106} {\bibfield  {journal} {\bibinfo  {journal} {J. High Energ. Phys.}\ }\textbf {\bibinfo {volume} {2016}},\ \bibinfo {pages} {106} (\bibinfo {year} {2016})}\BibitemShut {NoStop}%
\bibitem [{\citenamefont {Nahum}\ \emph {et~al.}(2018)\citenamefont {Nahum}, \citenamefont {Vijay},\ and\ \citenamefont {Haah}}]{nahum_operator_2018}%
  \BibitemOpen
  \bibfield  {author} {\bibinfo {author} {\bibfnamefont {A.}~\bibnamefont {Nahum}}, \bibinfo {author} {\bibfnamefont {S.}~\bibnamefont {Vijay}}, \ and\ \bibinfo {author} {\bibfnamefont {J.}~\bibnamefont {Haah}},\ }\href {https://link.aps.org/doi/10.1103/PhysRevX.8.021014} {\bibfield  {journal} {\bibinfo  {journal} {Phys. Rev. X}\ }\textbf {\bibinfo {volume} {8}},\ \bibinfo {pages} {021014} (\bibinfo {year} {2018})}\BibitemShut {NoStop}%
\bibitem [{\citenamefont {von Keyserlingk}\ \emph {et~al.}(2018)\citenamefont {von Keyserlingk}, \citenamefont {Rakovszky}, \citenamefont {Pollmann},\ and\ \citenamefont {Sondhi}}]{von_keyserlingk_operator_2018}%
  \BibitemOpen
  \bibfield  {author} {\bibinfo {author} {\bibfnamefont {C.~W.}\ \bibnamefont {von Keyserlingk}}, \bibinfo {author} {\bibfnamefont {T.}~\bibnamefont {Rakovszky}}, \bibinfo {author} {\bibfnamefont {F.}~\bibnamefont {Pollmann}}, \ and\ \bibinfo {author} {\bibfnamefont {S.~L.}\ \bibnamefont {Sondhi}},\ }\href {https://link.aps.org/doi/10.1103/PhysRevX.8.021013} {\bibfield  {journal} {\bibinfo  {journal} {Phys. Rev. X}\ }\textbf {\bibinfo {volume} {8}},\ \bibinfo {pages} {021013} (\bibinfo {year} {2018})}\BibitemShut {NoStop}%
\bibitem [{\citenamefont {Zhou}\ \emph {et~al.}(2020)\citenamefont {Zhou}, \citenamefont {Xu}, \citenamefont {Chen}, \citenamefont {Guo},\ and\ \citenamefont {Swingle}}]{zhou_operator_2020}%
  \BibitemOpen
  \bibfield  {author} {\bibinfo {author} {\bibfnamefont {T.}~\bibnamefont {Zhou}}, \bibinfo {author} {\bibfnamefont {S.}~\bibnamefont {Xu}}, \bibinfo {author} {\bibfnamefont {X.}~\bibnamefont {Chen}}, \bibinfo {author} {\bibfnamefont {A.}~\bibnamefont {Guo}}, \ and\ \bibinfo {author} {\bibfnamefont {B.}~\bibnamefont {Swingle}},\ }\href {\doibase 10.1103/physrevlett.124.180601} {\bibfield  {journal} {\bibinfo  {journal} {Phys. Rev. Lett.}\ }\textbf {\bibinfo {volume} {124}},\ \bibinfo {pages} {180601} (\bibinfo {year} {2020})}\BibitemShut {NoStop}%
\bibitem [{\citenamefont {Xu}\ and\ \citenamefont {Swingle}(2019{\natexlab{a}})}]{xu_locality_2018}%
  \BibitemOpen
  \bibfield  {author} {\bibinfo {author} {\bibfnamefont {S.}~\bibnamefont {Xu}}\ and\ \bibinfo {author} {\bibfnamefont {B.}~\bibnamefont {Swingle}},\ }\href {\doibase 10.1103/physrevx.9.031048} {\bibfield  {journal} {\bibinfo  {journal} {Phys. Rev. X}\ }\textbf {\bibinfo {volume} {9}} (\bibinfo {year} {2019}{\natexlab{a}}),\ 10.1103/physrevx.9.031048},\ \bibinfo {note} {arXiv: 1805.05376}\BibitemShut {NoStop}%
\bibitem [{\citenamefont {Chen}\ and\ \citenamefont {Zhou}(2019)}]{chen_quantum_2018}%
  \BibitemOpen
  \bibfield  {author} {\bibinfo {author} {\bibfnamefont {X.}~\bibnamefont {Chen}}\ and\ \bibinfo {author} {\bibfnamefont {T.}~\bibnamefont {Zhou}},\ }\href {\doibase 10.1103/physrevb.100.064305} {\bibfield  {journal} {\bibinfo  {journal} {Phys. Rev. B}\ }\textbf {\bibinfo {volume} {100}},\ \bibinfo {pages} {064305} (\bibinfo {year} {2019})}\BibitemShut {NoStop}%
\bibitem [{\citenamefont {Zhou}\ and\ \citenamefont {Chen}(2019)}]{zhou_operator_2018}%
  \BibitemOpen
  \bibfield  {author} {\bibinfo {author} {\bibfnamefont {T.}~\bibnamefont {Zhou}}\ and\ \bibinfo {author} {\bibfnamefont {X.}~\bibnamefont {Chen}},\ }\href {\doibase 10.1103/physreve.99.052212} {\bibfield  {journal} {\bibinfo  {journal} {Phys. Rev. E}\ }\textbf {\bibinfo {volume} {99}},\ \bibinfo {pages} {052212} (\bibinfo {year} {2019})}\BibitemShut {NoStop}%
\bibitem [{\citenamefont {Roberts}\ \emph {et~al.}(2018)\citenamefont {Roberts}, \citenamefont {Stanford},\ and\ \citenamefont {Streicher}}]{roberts_operator_2018}%
  \BibitemOpen
  \bibfield  {author} {\bibinfo {author} {\bibfnamefont {D.~A.}\ \bibnamefont {Roberts}}, \bibinfo {author} {\bibfnamefont {D.}~\bibnamefont {Stanford}}, \ and\ \bibinfo {author} {\bibfnamefont {A.}~\bibnamefont {Streicher}},\ }\href {\doibase 10.1007/jhep06(2018)122} {\bibfield  {journal} {\bibinfo  {journal} {J. High Energ. Phys.}\ }\textbf {\bibinfo {volume} {2018}},\ \bibinfo {pages} {122} (\bibinfo {year} {2018})}\BibitemShut {NoStop}%
\bibitem [{\citenamefont {Qi}\ and\ \citenamefont {Streicher}(2019)}]{qi_quantum_2018}%
  \BibitemOpen
  \bibfield  {author} {\bibinfo {author} {\bibfnamefont {X.-L.}\ \bibnamefont {Qi}}\ and\ \bibinfo {author} {\bibfnamefont {A.}~\bibnamefont {Streicher}},\ }\href {\doibase 10.1007/jhep08(2019)012} {\bibfield  {journal} {\bibinfo  {journal} {J. High Energ. Phys.}\ }\textbf {\bibinfo {volume} {2019}} (\bibinfo {year} {2019}),\ 10.1007/jhep08(2019)012},\ \Eprint {http://arxiv.org/abs/1810.11958} {arXiv:1810.11958 [cond-mat, physics:hep-th, physics:quant-ph]} \BibitemShut {NoStop}%
\bibitem [{\citenamefont {De}\ \emph {et~al.}(2024)\citenamefont {De}, \citenamefont {Borla}, \citenamefont {Cao},\ and\ \citenamefont {Gazit}}]{de_stochastic_2024}%
  \BibitemOpen
  \bibfield  {author} {\bibinfo {author} {\bibfnamefont {A.}~\bibnamefont {De}}, \bibinfo {author} {\bibfnamefont {U.}~\bibnamefont {Borla}}, \bibinfo {author} {\bibfnamefont {X.}~\bibnamefont {Cao}}, \ and\ \bibinfo {author} {\bibfnamefont {S.}~\bibnamefont {Gazit}},\ }\href {\doibase 10.1103/PhysRevB.110.155135} {\bibfield  {journal} {\bibinfo  {journal} {Physical Review B}\ }\textbf {\bibinfo {volume} {110}},\ \bibinfo {pages} {155135} (\bibinfo {year} {2024})},\ \Eprint {http://arxiv.org/abs/2401.06215} {arXiv:2401.06215 [cond-mat]} \BibitemShut {NoStop}%
\bibitem [{\citenamefont {Lashkari}\ \emph {et~al.}(2013)\citenamefont {Lashkari}, \citenamefont {Stanford}, \citenamefont {Hastings}, \citenamefont {Osborne},\ and\ \citenamefont {Hayden}}]{lashkari_towards_2013}%
  \BibitemOpen
  \bibfield  {author} {\bibinfo {author} {\bibfnamefont {N.}~\bibnamefont {Lashkari}}, \bibinfo {author} {\bibfnamefont {D.}~\bibnamefont {Stanford}}, \bibinfo {author} {\bibfnamefont {M.}~\bibnamefont {Hastings}}, \bibinfo {author} {\bibfnamefont {T.}~\bibnamefont {Osborne}}, \ and\ \bibinfo {author} {\bibfnamefont {P.}~\bibnamefont {Hayden}},\ }\href {\doibase 10.1007/jhep04(2013)022} {\bibfield  {journal} {\bibinfo  {journal} {J. High Energ. Phys.}\ }\textbf {\bibinfo {volume} {2013}} (\bibinfo {year} {2013}),\ 10.1007/jhep04(2013)022}\BibitemShut {NoStop}%
\bibitem [{\citenamefont {Agarwal}\ and\ \citenamefont {Xu}(2021)}]{agarwal_emergent_2021}%
  \BibitemOpen
  \bibfield  {author} {\bibinfo {author} {\bibfnamefont {L.}~\bibnamefont {Agarwal}}\ and\ \bibinfo {author} {\bibfnamefont {S.}~\bibnamefont {Xu}},\ }\href@noop {} {\bibfield  {journal} {\bibinfo  {journal} {arXiv:2108.05810 [cond-mat, physics:hep-th, physics:quant-ph]}\ } (\bibinfo {year} {2021})},\ \Eprint {http://arxiv.org/abs/2108.05810} {arXiv:2108.05810 [cond-mat, physics:hep-th, physics:quant-ph]} \BibitemShut {NoStop}%
\bibitem [{\citenamefont {Erd{\H o}s}\ and\ \citenamefont {Schr{\"o}der}(2014)}]{erdos_phase_2014}%
  \BibitemOpen
  \bibfield  {author} {\bibinfo {author} {\bibfnamefont {L.}~\bibnamefont {Erd{\H o}s}}\ and\ \bibinfo {author} {\bibfnamefont {D.}~\bibnamefont {Schr{\"o}der}},\ }\href {\doibase 10.1007/s11040-014-9164-3} {\bibfield  {journal} {\bibinfo  {journal} {Mathematical Physics, Analysis and Geometry}\ }\textbf {\bibinfo {volume} {17}},\ \bibinfo {pages} {441} (\bibinfo {year} {2014})},\ \Eprint {http://arxiv.org/abs/1407.1552} {arXiv:1407.1552} \BibitemShut {NoStop}%
\bibitem [{\citenamefont {Lucas}(2019)}]{lucas_quantum_2019}%
  \BibitemOpen
  \bibfield  {author} {\bibinfo {author} {\bibfnamefont {A.}~\bibnamefont {Lucas}},\ }\href@noop {} {\bibfield  {journal} {\bibinfo  {journal} {arXiv:1903.01468 [cond-mat, physics:hep-th, physics:quant-ph]}\ } (\bibinfo {year} {2019})},\ \Eprint {http://arxiv.org/abs/1903.01468} {arXiv:1903.01468 [cond-mat, physics:hep-th, physics:quant-ph]} \BibitemShut {NoStop}%
\bibitem [{\citenamefont {Vardhan}\ and\ \citenamefont {Moudgalya}(2024)}]{vardhan_entanglement_2024}%
  \BibitemOpen
  \bibfield  {author} {\bibinfo {author} {\bibfnamefont {S.}~\bibnamefont {Vardhan}}\ and\ \bibinfo {author} {\bibfnamefont {S.}~\bibnamefont {Moudgalya}},\ }\href {\doibase 10.48550/arXiv.2407.16763} {\enquote {\bibinfo {title} {Entanglement dynamics from universal low-lying modes},}\ } (\bibinfo {year} {2024}),\ \Eprint {http://arxiv.org/abs/2407.16763} {arXiv:2407.16763 [cond-mat, physics:hep-th, physics:quant-ph]} \BibitemShut {NoStop}%
\bibitem [{\citenamefont {Aleiner}\ \emph {et~al.}(2016)\citenamefont {Aleiner}, \citenamefont {Faoro},\ and\ \citenamefont {Ioffe}}]{aleiner_microscopic_2016}%
  \BibitemOpen
  \bibfield  {author} {\bibinfo {author} {\bibfnamefont {I.~L.}\ \bibnamefont {Aleiner}}, \bibinfo {author} {\bibfnamefont {L.}~\bibnamefont {Faoro}}, \ and\ \bibinfo {author} {\bibfnamefont {L.~B.}\ \bibnamefont {Ioffe}},\ }\href {\doibase 10.1016/j.aop.2016.09.006} {\bibfield  {journal} {\bibinfo  {journal} {Ann. Phys-new. York.}\ }\textbf {\bibinfo {volume} {375}},\ \bibinfo {pages} {378} (\bibinfo {year} {2016})},\ \Eprint {http://arxiv.org/abs/1609.01251} {arXiv:1609.01251 [cond-mat, physics:hep-th, physics:quant-ph]} \BibitemShut {NoStop}%
\bibitem [{\citenamefont {Aron}\ \emph {et~al.}(2023{\natexlab{a}})\citenamefont {Aron}, \citenamefont {Brunet},\ and\ \citenamefont {Mitra}}]{aron_kinetics_2023}%
  \BibitemOpen
  \bibfield  {author} {\bibinfo {author} {\bibfnamefont {C.}~\bibnamefont {Aron}}, \bibinfo {author} {\bibfnamefont {{\'E}.}~\bibnamefont {Brunet}}, \ and\ \bibinfo {author} {\bibfnamefont {A.}~\bibnamefont {Mitra}},\ }\href {\doibase 10.1103/PhysRevB.108.L241106} {\bibfield  {journal} {\bibinfo  {journal} {Physical Review B}\ }\textbf {\bibinfo {volume} {108}},\ \bibinfo {pages} {L241106} (\bibinfo {year} {2023}{\natexlab{a}})},\ \Eprint {http://arxiv.org/abs/2305.04958} {arXiv:2305.04958 [cond-mat, physics:quant-ph]} \BibitemShut {NoStop}%
\bibitem [{\citenamefont {Aron}\ \emph {et~al.}(2023{\natexlab{b}})\citenamefont {Aron}, \citenamefont {Brunet},\ and\ \citenamefont {Mitra}}]{aron_traveling_2023}%
  \BibitemOpen
  \bibfield  {author} {\bibinfo {author} {\bibfnamefont {C.}~\bibnamefont {Aron}}, \bibinfo {author} {\bibfnamefont {E.}~\bibnamefont {Brunet}}, \ and\ \bibinfo {author} {\bibfnamefont {A.}~\bibnamefont {Mitra}},\ }\href {\doibase 10.21468/SciPostPhys.15.2.042} {\bibfield  {journal} {\bibinfo  {journal} {SciPost Physics}\ }\textbf {\bibinfo {volume} {15}},\ \bibinfo {pages} {042} (\bibinfo {year} {2023}{\natexlab{b}})},\ \Eprint {http://arxiv.org/abs/2212.13265} {arXiv:2212.13265 [cond-mat, physics:quant-ph]} \BibitemShut {NoStop}%
\bibitem [{\citenamefont {Kolmogorov}\ \emph {et~al.}(1937)\citenamefont {Kolmogorov}, \citenamefont {Petrovsky},\ and\ \citenamefont {Piskunov}}]{kolmogorov_investigation_1937}%
  \BibitemOpen
  \bibfield  {author} {\bibinfo {author} {\bibfnamefont {A.}~\bibnamefont {Kolmogorov}}, \bibinfo {author} {\bibfnamefont {I.}~\bibnamefont {Petrovsky}}, \ and\ \bibinfo {author} {\bibfnamefont {N.}~\bibnamefont {Piskunov}},\ }\href@noop {} {\bibfield  {journal} {\bibinfo  {journal} {Bulletin of Moscow State University Series A: Mathematics and Mechanics}\ ,\ \bibinfo {pages} {1}} (\bibinfo {year} {1937})}\BibitemShut {NoStop}%
\bibitem [{\citenamefont {Fisher}(1937)}]{fisher_wave_1937}%
  \BibitemOpen
  \bibfield  {author} {\bibinfo {author} {\bibfnamefont {R.~A.}\ \bibnamefont {Fisher}},\ }\href {\doibase 10.1111/j.1469-1809.1937.tb02153.x} {\bibfield  {journal} {\bibinfo  {journal} {Ann. Eugenic.}\ }\textbf {\bibinfo {volume} {7}},\ \bibinfo {pages} {355} (\bibinfo {year} {1937})},\ \bibinfo {note} {\_eprint: https://onlinelibrary.wiley.com/doi/pdf/10.1111/j.1469-1809.1937.tb02153.x}\BibitemShut {NoStop}%
\bibitem [{\citenamefont {Ablowitz}\ and\ \citenamefont {Zeppetella}(1979)}]{ablowitz_explicit_1979}%
  \BibitemOpen
  \bibfield  {author} {\bibinfo {author} {\bibfnamefont {M.}~\bibnamefont {Ablowitz}}\ and\ \bibinfo {author} {\bibfnamefont {A.}~\bibnamefont {Zeppetella}},\ }\href {\doibase 10.1016/s0092-8240(79)80020-8} {\bibfield  {journal} {\bibinfo  {journal} {B. Math. Biol.}\ }\textbf {\bibinfo {volume} {41}},\ \bibinfo {pages} {835} (\bibinfo {year} {1979})}\BibitemShut {NoStop}%
\bibitem [{\citenamefont {Brunet}(2016)}]{brunet_aspects_2016}%
  \BibitemOpen
  \bibfield  {author} {\bibinfo {author} {\bibfnamefont {{\'E}.}~\bibnamefont {Brunet}}\ }(\bibinfo {year} {2016})\BibitemShut {NoStop}%
\bibitem [{\citenamefont {Xu}\ and\ \citenamefont {Swingle}(2019{\natexlab{b}})}]{Xu_Swingle_2018}%
  \BibitemOpen
  \bibfield  {author} {\bibinfo {author} {\bibfnamefont {S.}~\bibnamefont {Xu}}\ and\ \bibinfo {author} {\bibfnamefont {B.}~\bibnamefont {Swingle}},\ }\href {\doibase 10.1038/s41567-019-0712-4} {\bibfield  {journal} {\bibinfo  {journal} {Nat. Phys.}\ }\textbf {\bibinfo {volume} {16}},\ \bibinfo {eid} {arXiv:1802.00801} (\bibinfo {year} {2019}{\natexlab{b}})},\ \Eprint {http://arxiv.org/abs/1802.00801} {arXiv:1802.00801 [quant-ph]} \BibitemShut {NoStop}%
\bibitem [{\citenamefont {Bentsen}\ \emph {et~al.}(2021)\citenamefont {Bentsen}, \citenamefont {Sahu},\ and\ \citenamefont {Swingle}}]{bentsen_measurement-induced_2021}%
  \BibitemOpen
  \bibfield  {author} {\bibinfo {author} {\bibfnamefont {G.}~\bibnamefont {Bentsen}}, \bibinfo {author} {\bibfnamefont {S.}~\bibnamefont {Sahu}}, \ and\ \bibinfo {author} {\bibfnamefont {B.}~\bibnamefont {Swingle}},\ }\href@noop {} {\bibfield  {journal} {\bibinfo  {journal} {arXiv:2104.07688 [cond-mat, physics:hep-th, physics:quant-ph]}\ } (\bibinfo {year} {2021})},\ \Eprint {http://arxiv.org/abs/2104.07688} {arXiv:2104.07688 [cond-mat, physics:hep-th, physics:quant-ph]} \BibitemShut {NoStop}%
\bibitem [{\citenamefont {Roberts}\ and\ \citenamefont {Swingle}(2016)}]{roberts_lieb-robinson_2016}%
  \BibitemOpen
  \bibfield  {author} {\bibinfo {author} {\bibfnamefont {D.~A.}\ \bibnamefont {Roberts}}\ and\ \bibinfo {author} {\bibfnamefont {B.}~\bibnamefont {Swingle}},\ }\href {\doibase 10.1103/physrevlett.117.091602} {\bibfield  {journal} {\bibinfo  {journal} {Phys. Rev. Lett.}\ }\textbf {\bibinfo {volume} {117}},\ \bibinfo {pages} {091602} (\bibinfo {year} {2016})}\BibitemShut {NoStop}%
\bibitem [{\citenamefont {Doering}\ \emph {et~al.}(2003)\citenamefont {Doering}, \citenamefont {Mueller},\ and\ \citenamefont {Smereka}}]{doering_interacting_2003}%
  \BibitemOpen
  \bibfield  {author} {\bibinfo {author} {\bibfnamefont {C.~R.}\ \bibnamefont {Doering}}, \bibinfo {author} {\bibfnamefont {C.}~\bibnamefont {Mueller}}, \ and\ \bibinfo {author} {\bibfnamefont {P.}~\bibnamefont {Smereka}},\ }\href {\doibase 10.1016/S0378-4371(03)00203-6} {\bibfield  {journal} {\bibinfo  {journal} {Physica A: Statistical Mechanics and its Applications}\ }\bibinfo {series} {Stochastic {{Systems}}: {{From Randomness}} to {{Complexity}}},\ \textbf {\bibinfo {volume} {325}},\ \bibinfo {pages} {243} (\bibinfo {year} {2003})}\BibitemShut {NoStop}%
\bibitem [{\citenamefont {Hallatschek}\ and\ \citenamefont {Fisher}(2014)}]{hallatschek_acceleration_2014}%
  \BibitemOpen
  \bibfield  {author} {\bibinfo {author} {\bibfnamefont {O.}~\bibnamefont {Hallatschek}}\ and\ \bibinfo {author} {\bibfnamefont {D.~S.}\ \bibnamefont {Fisher}},\ }\href {\doibase 10.1073/pnas.1404663111} {\bibfield  {journal} {\bibinfo  {journal} {Proc. Natl. Acad. Sci. U.S.A.}\ }\textbf {\bibinfo {volume} {111}},\ \bibinfo {pages} {E4911} (\bibinfo {year} {2014})}\BibitemShut {NoStop}%
\bibitem [{SM()}]{SM}%
  \BibitemOpen
  \href@noop {} {}\bibinfo {note} {See Supplemental Material for an overview of the algorithm, estimation of $\beta$ given equilibrium height, derivation of the duality relation and the scaling function close to $\alpha = 1$.}\BibitemShut {Stop}%
\bibitem [{\citenamefont {Chatterjee}\ and\ \citenamefont {S.~Dey}(2015)}]{chatterjee_multiple_2013}%
  \BibitemOpen
  \bibfield  {author} {\bibinfo {author} {\bibfnamefont {S.}~\bibnamefont {Chatterjee}}\ and\ \bibinfo {author} {\bibfnamefont {P.}~\bibnamefont {S.~Dey}},\ }\href {\doibase 10.1002/cpa.21571} {\bibfield  {journal} {\bibinfo  {journal} {Commun. Pur. Appl. Math.}\ }\textbf {\bibinfo {volume} {69}},\ \bibinfo {pages} {203} (\bibinfo {year} {2015})}\BibitemShut {NoStop}%
\bibitem [{\citenamefont {Zhou}\ \emph {et~al.}(2023)\citenamefont {Zhou}, \citenamefont {Guo}, \citenamefont {Xu}, \citenamefont {Chen},\ and\ \citenamefont {Swingle}}]{zhou_hydrodynamic_2023}%
  \BibitemOpen
  \bibfield  {author} {\bibinfo {author} {\bibfnamefont {T.}~\bibnamefont {Zhou}}, \bibinfo {author} {\bibfnamefont {A.~Y.}\ \bibnamefont {Guo}}, \bibinfo {author} {\bibfnamefont {S.}~\bibnamefont {Xu}}, \bibinfo {author} {\bibfnamefont {X.}~\bibnamefont {Chen}}, \ and\ \bibinfo {author} {\bibfnamefont {B.}~\bibnamefont {Swingle}},\ }\href {\doibase 10.1103/PhysRevB.107.014201} {\bibfield  {journal} {\bibinfo  {journal} {Physical Review B}\ }\textbf {\bibinfo {volume} {107}},\ \bibinfo {pages} {014201} (\bibinfo {year} {2023})},\ \Eprint {http://arxiv.org/abs/2208.01649} {arXiv:2208.01649 [cond-mat, physics:physics, physics:quant-ph]} \BibitemShut {NoStop}%
\bibitem [{\citenamefont {Álvarez}\ and\ \citenamefont {Suter}(2011)}]{alvarez_localization_2011}%
  \BibitemOpen
  \bibfield  {author} {\bibinfo {author} {\bibfnamefont {G.~A.}\ \bibnamefont {Álvarez}}\ and\ \bibinfo {author} {\bibfnamefont {D.}~\bibnamefont {Suter}},\ }\href {\doibase 10.1103/physreva.84.012320} {\bibfield  {journal} {\bibinfo  {journal} {Phys. Rev. A}\ }\textbf {\bibinfo {volume} {84}},\ \bibinfo {pages} {012320} (\bibinfo {year} {2011})}\BibitemShut {NoStop}%
\bibitem [{\citenamefont {Mi}\ \emph {et~al.}(2021)\citenamefont {Mi}, \citenamefont {Roushan}, \citenamefont {Quintana}, \citenamefont {Mandr{\`a}}, \citenamefont {Marshall}, \citenamefont {Neill}, \citenamefont {Arute}, \citenamefont {Arya}, \citenamefont {Atalaya}, \citenamefont {Babbush}, \citenamefont {Bardin}, \citenamefont {Barends}, \citenamefont {Basso}, \citenamefont {Bengtsson}, \citenamefont {Boixo}, \citenamefont {Bourassa}, \citenamefont {Broughton}, \citenamefont {Buckley}, \citenamefont {Buell}, \citenamefont {Burkett}, \citenamefont {Bushnell}, \citenamefont {Chen}, \citenamefont {Chiaro}, \citenamefont {Collins}, \citenamefont {Courtney}, \citenamefont {Demura}, \citenamefont {Derk}, \citenamefont {Dunsworth}, \citenamefont {Eppens}, \citenamefont {Erickson}, \citenamefont {Farhi}, \citenamefont {Fowler}, \citenamefont {Foxen}, \citenamefont {Gidney}, \citenamefont {Giustina}, \citenamefont {Gross}, \citenamefont {Harrigan}, \citenamefont {Harrington}, \citenamefont {Hilton}, \citenamefont
  {Ho}, \citenamefont {Hong}, \citenamefont {Huang}, \citenamefont {Huggins}, \citenamefont {Ioffe}, \citenamefont {Isakov}, \citenamefont {Jeffrey}, \citenamefont {Jiang}, \citenamefont {Jones}, \citenamefont {Kafri}, \citenamefont {Kelly}, \citenamefont {Kim}, \citenamefont {Kitaev}, \citenamefont {Klimov}, \citenamefont {Korotkov}, \citenamefont {Kostritsa}, \citenamefont {Landhuis}, \citenamefont {Laptev}, \citenamefont {Lucero}, \citenamefont {Martin}, \citenamefont {McClean}, \citenamefont {McCourt}, \citenamefont {McEwen}, \citenamefont {Megrant}, \citenamefont {Miao}, \citenamefont {Mohseni}, \citenamefont {Montazeri}, \citenamefont {Mruczkiewicz}, \citenamefont {Mutus}, \citenamefont {Naaman}, \citenamefont {Neeley}, \citenamefont {Newman}, \citenamefont {Niu}, \citenamefont {O'Brien}, \citenamefont {Opremcak}, \citenamefont {Ostby}, \citenamefont {Pato}, \citenamefont {Petukhov}, \citenamefont {Redd}, \citenamefont {Rubin}, \citenamefont {Sank}, \citenamefont {Satzinger}, \citenamefont {Shvarts},
  \citenamefont {Strain}, \citenamefont {Szalay}, \citenamefont {Trevithick}, \citenamefont {Villalonga}, \citenamefont {White}, \citenamefont {Yao}, \citenamefont {Yeh}, \citenamefont {Zalcman}, \citenamefont {Neven}, \citenamefont {Aleiner}, \citenamefont {Kechedzhi}, \citenamefont {Smelyanskiy},\ and\ \citenamefont {Chen}}]{mi_information_2021}%
  \BibitemOpen
  \bibfield  {author} {\bibinfo {author} {\bibfnamefont {X.}~\bibnamefont {Mi}}, \bibinfo {author} {\bibfnamefont {P.}~\bibnamefont {Roushan}}, \bibinfo {author} {\bibfnamefont {C.}~\bibnamefont {Quintana}}, \bibinfo {author} {\bibfnamefont {S.}~\bibnamefont {Mandr{\`a}}}, \bibinfo {author} {\bibfnamefont {J.}~\bibnamefont {Marshall}}, \bibinfo {author} {\bibfnamefont {C.}~\bibnamefont {Neill}}, \bibinfo {author} {\bibfnamefont {F.}~\bibnamefont {Arute}}, \bibinfo {author} {\bibfnamefont {K.}~\bibnamefont {Arya}}, \bibinfo {author} {\bibfnamefont {J.}~\bibnamefont {Atalaya}}, \bibinfo {author} {\bibfnamefont {R.}~\bibnamefont {Babbush}}, \bibinfo {author} {\bibfnamefont {J.~C.}\ \bibnamefont {Bardin}}, \bibinfo {author} {\bibfnamefont {R.}~\bibnamefont {Barends}}, \bibinfo {author} {\bibfnamefont {J.}~\bibnamefont {Basso}}, \bibinfo {author} {\bibfnamefont {A.}~\bibnamefont {Bengtsson}}, \bibinfo {author} {\bibfnamefont {S.}~\bibnamefont {Boixo}}, \bibinfo {author} {\bibfnamefont {A.}~\bibnamefont
  {Bourassa}}, \bibinfo {author} {\bibfnamefont {M.}~\bibnamefont {Broughton}}, \bibinfo {author} {\bibfnamefont {B.~B.}\ \bibnamefont {Buckley}}, \bibinfo {author} {\bibfnamefont {D.~A.}\ \bibnamefont {Buell}}, \bibinfo {author} {\bibfnamefont {B.}~\bibnamefont {Burkett}}, \bibinfo {author} {\bibfnamefont {N.}~\bibnamefont {Bushnell}}, \bibinfo {author} {\bibfnamefont {Z.}~\bibnamefont {Chen}}, \bibinfo {author} {\bibfnamefont {B.}~\bibnamefont {Chiaro}}, \bibinfo {author} {\bibfnamefont {R.}~\bibnamefont {Collins}}, \bibinfo {author} {\bibfnamefont {W.}~\bibnamefont {Courtney}}, \bibinfo {author} {\bibfnamefont {S.}~\bibnamefont {Demura}}, \bibinfo {author} {\bibfnamefont {A.~R.}\ \bibnamefont {Derk}}, \bibinfo {author} {\bibfnamefont {A.}~\bibnamefont {Dunsworth}}, \bibinfo {author} {\bibfnamefont {D.}~\bibnamefont {Eppens}}, \bibinfo {author} {\bibfnamefont {C.}~\bibnamefont {Erickson}}, \bibinfo {author} {\bibfnamefont {E.}~\bibnamefont {Farhi}}, \bibinfo {author} {\bibfnamefont {A.~G.}\ \bibnamefont
  {Fowler}}, \bibinfo {author} {\bibfnamefont {B.}~\bibnamefont {Foxen}}, \bibinfo {author} {\bibfnamefont {C.}~\bibnamefont {Gidney}}, \bibinfo {author} {\bibfnamefont {M.}~\bibnamefont {Giustina}}, \bibinfo {author} {\bibfnamefont {J.~A.}\ \bibnamefont {Gross}}, \bibinfo {author} {\bibfnamefont {M.~P.}\ \bibnamefont {Harrigan}}, \bibinfo {author} {\bibfnamefont {S.~D.}\ \bibnamefont {Harrington}}, \bibinfo {author} {\bibfnamefont {J.}~\bibnamefont {Hilton}}, \bibinfo {author} {\bibfnamefont {A.}~\bibnamefont {Ho}}, \bibinfo {author} {\bibfnamefont {S.}~\bibnamefont {Hong}}, \bibinfo {author} {\bibfnamefont {T.}~\bibnamefont {Huang}}, \bibinfo {author} {\bibfnamefont {W.~J.}\ \bibnamefont {Huggins}}, \bibinfo {author} {\bibfnamefont {L.~B.}\ \bibnamefont {Ioffe}}, \bibinfo {author} {\bibfnamefont {S.~V.}\ \bibnamefont {Isakov}}, \bibinfo {author} {\bibfnamefont {E.}~\bibnamefont {Jeffrey}}, \bibinfo {author} {\bibfnamefont {Z.}~\bibnamefont {Jiang}}, \bibinfo {author} {\bibfnamefont {C.}~\bibnamefont
  {Jones}}, \bibinfo {author} {\bibfnamefont {D.}~\bibnamefont {Kafri}}, \bibinfo {author} {\bibfnamefont {J.}~\bibnamefont {Kelly}}, \bibinfo {author} {\bibfnamefont {S.}~\bibnamefont {Kim}}, \bibinfo {author} {\bibfnamefont {A.}~\bibnamefont {Kitaev}}, \bibinfo {author} {\bibfnamefont {P.~V.}\ \bibnamefont {Klimov}}, \bibinfo {author} {\bibfnamefont {A.~N.}\ \bibnamefont {Korotkov}}, \bibinfo {author} {\bibfnamefont {F.}~\bibnamefont {Kostritsa}}, \bibinfo {author} {\bibfnamefont {D.}~\bibnamefont {Landhuis}}, \bibinfo {author} {\bibfnamefont {P.}~\bibnamefont {Laptev}}, \bibinfo {author} {\bibfnamefont {E.}~\bibnamefont {Lucero}}, \bibinfo {author} {\bibfnamefont {O.}~\bibnamefont {Martin}}, \bibinfo {author} {\bibfnamefont {J.~R.}\ \bibnamefont {McClean}}, \bibinfo {author} {\bibfnamefont {T.}~\bibnamefont {McCourt}}, \bibinfo {author} {\bibfnamefont {M.}~\bibnamefont {McEwen}}, \bibinfo {author} {\bibfnamefont {A.}~\bibnamefont {Megrant}}, \bibinfo {author} {\bibfnamefont {K.~C.}\ \bibnamefont {Miao}},
  \bibinfo {author} {\bibfnamefont {M.}~\bibnamefont {Mohseni}}, \bibinfo {author} {\bibfnamefont {S.}~\bibnamefont {Montazeri}}, \bibinfo {author} {\bibfnamefont {W.}~\bibnamefont {Mruczkiewicz}}, \bibinfo {author} {\bibfnamefont {J.}~\bibnamefont {Mutus}}, \bibinfo {author} {\bibfnamefont {O.}~\bibnamefont {Naaman}}, \bibinfo {author} {\bibfnamefont {M.}~\bibnamefont {Neeley}}, \bibinfo {author} {\bibfnamefont {M.}~\bibnamefont {Newman}}, \bibinfo {author} {\bibfnamefont {M.~Y.}\ \bibnamefont {Niu}}, \bibinfo {author} {\bibfnamefont {T.~E.}\ \bibnamefont {O'Brien}}, \bibinfo {author} {\bibfnamefont {A.}~\bibnamefont {Opremcak}}, \bibinfo {author} {\bibfnamefont {E.}~\bibnamefont {Ostby}}, \bibinfo {author} {\bibfnamefont {B.}~\bibnamefont {Pato}}, \bibinfo {author} {\bibfnamefont {A.}~\bibnamefont {Petukhov}}, \bibinfo {author} {\bibfnamefont {N.}~\bibnamefont {Redd}}, \bibinfo {author} {\bibfnamefont {N.~C.}\ \bibnamefont {Rubin}}, \bibinfo {author} {\bibfnamefont {D.}~\bibnamefont {Sank}}, \bibinfo
  {author} {\bibfnamefont {K.~J.}\ \bibnamefont {Satzinger}}, \bibinfo {author} {\bibfnamefont {V.}~\bibnamefont {Shvarts}}, \bibinfo {author} {\bibfnamefont {D.}~\bibnamefont {Strain}}, \bibinfo {author} {\bibfnamefont {M.}~\bibnamefont {Szalay}}, \bibinfo {author} {\bibfnamefont {M.~D.}\ \bibnamefont {Trevithick}}, \bibinfo {author} {\bibfnamefont {B.}~\bibnamefont {Villalonga}}, \bibinfo {author} {\bibfnamefont {T.}~\bibnamefont {White}}, \bibinfo {author} {\bibfnamefont {Z.~J.}\ \bibnamefont {Yao}}, \bibinfo {author} {\bibfnamefont {P.}~\bibnamefont {Yeh}}, \bibinfo {author} {\bibfnamefont {A.}~\bibnamefont {Zalcman}}, \bibinfo {author} {\bibfnamefont {H.}~\bibnamefont {Neven}}, \bibinfo {author} {\bibfnamefont {I.}~\bibnamefont {Aleiner}}, \bibinfo {author} {\bibfnamefont {K.}~\bibnamefont {Kechedzhi}}, \bibinfo {author} {\bibfnamefont {V.}~\bibnamefont {Smelyanskiy}}, \ and\ \bibinfo {author} {\bibfnamefont {Y.}~\bibnamefont {Chen}},\ }\href {\doibase 10.1126/science.abg5029} {\bibfield  {journal}
  {\bibinfo  {journal} {Science}\ }\textbf {\bibinfo {volume} {374}},\ \bibinfo {pages} {1479} (\bibinfo {year} {2021})}\BibitemShut {NoStop}%
\bibitem [{\citenamefont {Jacoby}\ \emph {et~al.}(2024)\citenamefont {Jacoby}, \citenamefont {Huse},\ and\ \citenamefont {Gopalakrishnan}}]{jacoby_spectral_2024}%
  \BibitemOpen
  \bibfield  {author} {\bibinfo {author} {\bibfnamefont {J.~A.}\ \bibnamefont {Jacoby}}, \bibinfo {author} {\bibfnamefont {D.~A.}\ \bibnamefont {Huse}}, \ and\ \bibinfo {author} {\bibfnamefont {S.}~\bibnamefont {Gopalakrishnan}},\ }\href {\doibase 10.48550/arXiv.2409.17238} {\enquote {\bibinfo {title} {Spectral gaps of local quantum channels in the weak-dissipation limit},}\ } (\bibinfo {year} {2024}),\ \Eprint {http://arxiv.org/abs/2409.17238} {arXiv:2409.17238 [cond-mat, physics:quant-ph]} \BibitemShut {NoStop}%
\bibitem [{\citenamefont {O'Dea}\ \emph {et~al.}(2024)\citenamefont {O'Dea}, \citenamefont {Bhattacharjee}, \citenamefont {Gopalakrishnan},\ and\ \citenamefont {Khemani}}]{odea_absorbing_2024}%
  \BibitemOpen
  \bibfield  {author} {\bibinfo {author} {\bibfnamefont {N.}~\bibnamefont {O'Dea}}, \bibinfo {author} {\bibfnamefont {S.}~\bibnamefont {Bhattacharjee}}, \bibinfo {author} {\bibfnamefont {S.}~\bibnamefont {Gopalakrishnan}}, \ and\ \bibinfo {author} {\bibfnamefont {V.}~\bibnamefont {Khemani}},\ }\href@noop {} {\enquote {\bibinfo {title} {Absorbing state transitions with long-range annihilation},}\ } (\bibinfo {year} {2024})\BibitemShut {NoStop}%
\bibitem [{\citenamefont {Gillespie}(1976)}]{gillespie_general_1976}%
  \BibitemOpen
  \bibfield  {author} {\bibinfo {author} {\bibfnamefont {D.~T.}\ \bibnamefont {Gillespie}},\ }\href {\doibase 10.1016/0021-9991(76)90041-3} {\bibfield  {journal} {\bibinfo  {journal} {Journal of Computational Physics}\ }\textbf {\bibinfo {volume} {22}},\ \bibinfo {pages} {403} (\bibinfo {year} {1976})}\BibitemShut {NoStop}%
\bibitem [{\citenamefont {Gillespie}(1977)}]{gillespie_exact_1977}%
  \BibitemOpen
  \bibfield  {author} {\bibinfo {author} {\bibfnamefont {D.~T.}\ \bibnamefont {Gillespie}},\ }\href {\doibase 10.1021/j100540a008} {\bibfield  {journal} {\bibinfo  {journal} {The Journal of Physical Chemistry}\ }\textbf {\bibinfo {volume} {81}},\ \bibinfo {pages} {2340} (\bibinfo {year} {1977})}\BibitemShut {NoStop}%
\bibitem [{\citenamefont {Cao}\ \emph {et~al.}(2006)\citenamefont {Cao}, \citenamefont {Gillespie},\ and\ \citenamefont {Petzold}}]{cao_efficient_2006}%
  \BibitemOpen
  \bibfield  {author} {\bibinfo {author} {\bibfnamefont {Y.}~\bibnamefont {Cao}}, \bibinfo {author} {\bibfnamefont {D.~T.}\ \bibnamefont {Gillespie}}, \ and\ \bibinfo {author} {\bibfnamefont {L.~R.}\ \bibnamefont {Petzold}},\ }\href {\doibase 10.1063/1.2159468} {\bibfield  {journal} {\bibinfo  {journal} {The Journal of Chemical Physics}\ }\textbf {\bibinfo {volume} {124}},\ \bibinfo {pages} {044109} (\bibinfo {year} {2006})}\BibitemShut {NoStop}%
\bibitem [{\citenamefont {Gillespie}(2001)}]{gillespie_approximate_2001}%
  \BibitemOpen
  \bibfield  {author} {\bibinfo {author} {\bibfnamefont {D.~T.}\ \bibnamefont {Gillespie}},\ }\href {\doibase 10.1063/1.1378322} {\bibfield  {journal} {\bibinfo  {journal} {The Journal of Chemical Physics}\ }\textbf {\bibinfo {volume} {115}},\ \bibinfo {pages} {1716} (\bibinfo {year} {2001})}\BibitemShut {NoStop}%
\end{thebibliography}%
\let\addcontentsline\oldaddcontentsline

\end{document}